\documentclass[english,prd,nofootinbib,notitlepage,preprintnumbers]{revtex4-1}

\makeatletter

\usepackage[T1]{fontenc}
\usepackage[latin1]{inputenc}
\usepackage{amssymb,amsmath}
\usepackage{babel}
\usepackage{dcolumn,tabularx}
\usepackage{natbib}
\usepackage{url}

\usepackage{ifpdf}
\ifpdf   %if using pdfLaTeX in PDF mode
   \usepackage[pdftex]{graphicx}
   \DeclareGraphicsExtensions{.pdf,.png,.eps}
   \usepackage{pgf}
\else %if using LaTeX or pdfLaTeX in DVI mode
   \usepackage{graphics}
   \DeclareGraphicsExtensions{.eps,.png}   %\DeclareGraphicsExtensions{.eps,.bmp}
\fi

\usepackage[hyperfootnotes=false,pdfstartview=,colorlinks=true,linkcolor=purple,citecolor=blue,urlcolor=blue,bookmarks=false]{hyperref}   

\@ifundefined{definecolor}
 {\usepackage{color}}{}

\begin{document}

\title{Measuring our peculiar velocity on the CMB with high-multipole off-diagonal correlations}

\author{Luca Amendola$^{1}$}
%\email{l.amendola@thphys.uni-heidelberg.de}

\author{Riccardo Catena$^{1}$}
%\email{r.catena@thphys.uni-heidelberg.de}

\author{Isabella Masina$^{2,3}$}
%\email{masina@fe.infn.it}

\author{Alessio Notari$^{1,4}$}
%\email{a.notari@thphys.uni-heidelberg.de}

\author{Miguel Quartin$^{1,5}$}
%\email{mquartin@if.ufrj.br}

\author{Claudia Quercellini$^{6}$\vspace{.2cm}}
%\email{Claudia.Quercellini@uniroma2.it}

\affiliation{$^{1}$ Institute of Theoretical Physics, University of Heidelberg,
Philosophenweg 16, 69120 Heidelberg, Germany}

\affiliation{$^{2}$ Dip.~di Fisica, Università di Ferrara and INFN Sez.~di
Ferrara, Via Saragat 1, I-44100 Ferrara, Italy}

\affiliation{$^{3}$ CP3-Origins, IFK and IMADA, University of Southern Denmark, Campusvej
55, DK-5230, Odense M, Denmark}

\affiliation{$^{4}$ Departament de F\'isica Fonamental i Institut de Ci\`encies del Cosmos, Universitat de Barcelona, Mart\'i i Franqu\`es 1, 08028 Barcelona, Spain}

\affiliation{$^{5}$ Instituto de Física, Universidade Federal do Rio de Janeiro, CEP 21941-972, Rio de Janeiro, RJ, Brazil}

\affiliation{$^{6}$ Dip. di Fisica, Università di Roma \textquotedbl{}Tor Vergata\textquotedbl{},
Rome, Italy}

\begin{abstract}
Our peculiar velocity with respect to the CMB rest frame is known
to induce a large dipole in the CMB. However, the motion of an observer
has also the effect of distorting the anisotropies at all scales,
as shown by Challinor and Van Leeuwen (2002), due to aberration and
Doppler effects. We propose to measure independently our local motion
by using off-diagonal two-point correlation functions for high multipoles.
We study the observability of the signal for temperature and polarization
anisotropies. We point out that Planck can measure the velocity
$\beta$ with an error of about $30\%$ and the direction with an
error of about $20^\circ$. This method constitutes a cross-check,
which can be useful to verify that our CMB dipole is due mainly to
our velocity or to disentangle the velocity from other possible intrinsic
sources.

Although in this paper we focus on our peculiar velocity, a similar
effect would result also from other intrinsic vectorial distortion of
the CMB which would induce a dipolar lensing. Measuring
the off-diagonal correlation terms is therefore a test for a preferred
direction on the CMB sky.
\end{abstract}


\maketitle

\section{Introduction}

\global\long\def\dd{\textrm{d}}
\global\long\def\vp{\varphi}

\global\long\def\bI{\textrm{I}}
\global\long\def\n{{\bf {n}}}

Let us consider the motion of an observer (us) with respect to the cosmic microwave background (CMB) rest frame, with peculiar velocity ${\boldsymbol \beta}$. The motion of an observer has two effects: aberration, which is the apparent deflection of a light bundle due to the motion of the observer, and a Doppler effect on the frequency of the photons. The largest effect is due to a Doppler shift, which leads (to order $\beta\equiv| {\boldsymbol \beta}|$) to a large dipole in the CMB, even in a perfectly homogeneous sky. Because of the motion of the solar system barycenter with respect to the CMB rest frame, it is usually assumed
that the observed dipole entirely comes from such peculiar velocity effect, leading to the measurements $\beta=(1.231\pm0.003)\times10^{-3}$, $l=263.99^{\circ}\pm0.14^{\circ}$ and $b=48.26^{\circ}\pm0.03^{\circ}$
in galactic coordinates \cite{lineweaver96,WMAP1,WMAP2,PDG}.

However, it seems challenging to disentangle the effect of the local velocity from other possible sources, such as an intrinsic dipole or a secondary effect or a global dipolar anisotropy of the Universe. In particular, given the fact that the low multipoles appear to have alignments along some particular directions~\cite{Bennett:2010jb}, it would be interesting to check whether the dipole itself may contain anomalous contributions in addition to the Doppler effect. For these reasons, it would be useful to have an independent measurement of our peculiar velocity with respect to the CMB rest frame.

We propose to do this using the fact that aberration and Doppler effect distort the primordial anisotropies and introduce correlations among different multipoles: for the off-diagonal components (different $\ell$'s) the correlations are already at first order in $\beta$; for the diagonal ones (same $\ell$'s) they are only of second order~\cite{challinor02} (although~\cite{pereira10} claims that using a cut-sky as opposed to full-sky might lead to first order corrections on the diagonal terms). Our main aim in this paper is to show that the off-diagonal two-point correlation functions can be used by high-resolution experiments, such as the Planck satellite, to \textit{measure} our peculiar velocity, even without using information from the dipole.

In the previous literature only Ref.~\cite{burles06} has shown that peculiar velocities could be measured using the asymmetry in the location of the peaks of the power spectrum between forward and backward hemispheres, achieving a possible detection of $\beta$ at $2-3\sigma$ for resolution $\ell=1000-1500$. In~\cite{challinor02} the authors have computed the effect on the $a_{\ell m}$, in order to show that the contamination on cosmological parameter estimation due to our velocity is negligible. Finally~\cite{Menzies} has shown how to remove both the aberration and Doppler effects from CMB sky maps: by assuming that all of the CMB dipole is due to the relative motion between the instrument and the CMB, one can use this fiducial value of our peculiar velocity to ``de-boost'' the data, a procedure that should be preferentially  carried out in the raw measured time-ordered data.

Note that an effect similar to the one we discuss here could be generated also by other effects, such as a global dipolar anisotropy of the Universe, provided they induce a large dipolar lensing effect on the CMB photons. Generally speaking, therefore, the detection of such a correlation can be a measure of the existence of a preferred direction in our Universe.

%%%%%%%%%%%%%%%%%%%%%%%%%%%%%%%%%%%%%%%%%%%%%%%%%%%%%

\section{Aberration and Doppler effect}

If $\mathbf{\hat{n}}$ is the direction of the incoming light in a reference frame $S$ at rest with respect to the CMB frame, the direction $\mathbf{\hat{n}}^{\prime}$ observed (for the same event) in a reference frame $S^{\prime}$ which moves with a velocity $\mathbf{v}$ relative to $S$ can be calculated by applying a Lorentz transformation to the vector $\mathbf{V}=-\hat{\mathbf{n}}$, which locally describes the velocity of the light bundle. If we denote by $\mathbf{V}'=-\hat{\mathbf{n}}^{\prime}$ the velocity in the boosted frame, then the Lorentz transformations we are interested in are (we follow a derivation similar to the one in~\cite{challinor02})
\begin{equation}
V_{u}^{\prime}=\frac{V_{u}-\beta}{1-\beta V_{u}}\,;\qquad\qquad V_{w}^{\prime}=\frac{V_{w}}{\gamma(1-\beta V_{u})}\,,\end{equation}
 where $|\mathbf{v}|=\beta$ and $\gamma^{-2}=1-\beta^{2}$. The two
vectors $\hat{\mathbf{u}}$ and $\hat{\mathbf{w}}$ are parallel and
orthogonal to $\mathbf{v}$ respectively. They are such that $V_{u}=\hat{\mathbf{u}}\cdot\mathbf{V}=-\hat{\mathbf{u}}\cdot\mathbf{n}=-\cos\theta$
and $V_{w}=\hat{\mathbf{w}}\cdot\mathbf{V}=-\hat{\mathbf{w}}\cdot\mathbf{n}=-\sin\theta$,
where $\theta$ is the angle between $\mathbf{v}$ and $\hat{\mathbf{n}}$. With
this notation we can now explicitly write the vector $\hat{\mathbf{n}}^{\prime}$
as follows \begin{eqnarray}
\hat{\mathbf{n}}^{\prime} & = & -V_{u}^{\prime}\,\hat{\mathbf{u}}-V_{w}^{\prime}\,\hat{\mathbf{w}} \nonumber\\
 & = & -\frac{V_{u}-\beta}{1-\beta V_{u}}\,\hat{\mathbf{u}}-\frac{V_{w}}{\gamma(1-\beta V_{u})}\frac{\left[\hat{\mathbf{n}}-(\hat{\mathbf{n}}\cdot\hat{\mathbf{u}})\hat{\mathbf{u}}\right]}{\sqrt{1-(\hat{\mathbf{n}}\cdot\hat{\mathbf{u}})^{2}}} \nonumber\\
 & = & \frac{\hat{\mathbf{n}}\cdot\hat{\mathbf{u}}+\beta}{1+\beta\hat{\mathbf{n}}\cdot\hat{\mathbf{u}}}\,\hat{\mathbf{u}}+\frac{\left[\hat{\mathbf{n}}-(\hat{\mathbf{n}}\cdot\hat{\mathbf{u}})\hat{\mathbf{u}}\right]}{\gamma(1+\beta\hat{\mathbf{n}}\cdot\hat{\mathbf{u}})}\,.\end{eqnarray}

Let us choose now spherical coordinates, using $\hat{\mathbf{u}}$ as the $\hat{\mathbf{z}}$ axis, and $\theta$, $\varphi$ as the
usual spherical coordinates for the frame $S$. We do the same for the frame $S'$, calling $\theta'$ the angle between $\mathbf{v}$ and $\hat{\mathbf{n}}'$. The observer $S'$ finds a deviation in the direction of arrival of the photons with respect to $S$. The aberration angle is defined as ${\boldsymbol \alpha}\equiv  \hat{\mathbf{n'}}-\hat{\mathbf{n}}$. Its projection along the $\hat{\mathbf{z}}$ axis is given by:
% \theta-\theta'$
\begin{equation}\label{eq:alpha}
    {\boldsymbol \alpha} \cdot \hat{\mathbf{z}} \;= \frac{\beta \sin^2\theta }{1+\beta  \cos\theta} \, ,
\end{equation}
which means that at lowest order in $\beta$ we simply have:
%and is given by:
\begin{equation}
%\label{eq:alpha}
    \theta-\theta'
  %  \alpha\;=\;\theta -\arccos \left[\frac{\beta +\cos\theta }{1+\beta  \cos\theta}\right]
    \;=\; \beta~\sin\theta + \mathcal{O}(\beta ^2)\,.
\end{equation}

The peculiar motion of an observer with respect to a given source induces also a change in the frequency of the observed photons. According to Special Relativity the frequency $\nu^{\prime}$ in the boosted frame is related to the frequency $\nu$ measured at rest by the following relation
\begin{equation}
\nu^{\prime}=\nu\gamma(1+\beta\hat{\mathbf{n}}\cdot\hat{\mathbf{u}})\,.
\end{equation}
This is the usual Doppler effect.

%%%%%%%%%%%%%%%%%%%%%%%%%%%%%%%%%%%%%%%%%%%%%%%%%%%%%%%%%%%%%

\section{Effect on the multipoles}
\label{sec:alm}
It is instructive to consider, first, only the effect of the change of direction - \textit{i.e.} the aberration - on the CMB.  This effect can be treated with the same formalism which is used for CMB lensing so we will sometimes use the term ``lensing'' to refer to the effect of aberration. In fact, both effects amount to a redefinition of the direction of arrival of the primordial photons. In keeping with the standard early universe cosmology, we assume the primordial fluctuations to be Gaussian. It is also clear from this treatment that any large scale dipolar gravitational potential would induce a similar effect via lensing.

The observed temperature fluctuation $\Delta\equiv\Delta T/T$
due to lensing is usually computed in a gradient expansion~\cite{Bernardeau,Lewis} as a function of the unlensed fluctuation (or primordial, superscript $P$), where the total fluctuation is given by
\begin{equation}
\Delta({\bf \hat{n}'})\sim\Delta^{(P)}({\bf \hat{n}})+\Delta_{;i}^{(P)}({\bf \hat{n}})\Theta^{;i}({\bf \hat{n}})+\Delta_{;ij}^{(P)}({\bf \hat{n}})\Theta^{;j}({\bf \hat{n}})\Theta^{;i}({\bf \hat{n}})+\,...\,,\label{templensing}
\end{equation}
where the semicolon stands for the covariant derivative (in the 2-sphere) with respect to the $i$-th coordinate and here $\nabla\Theta\equiv {\boldsymbol \alpha}$. Here and in the following we adopt the exact all-sky formalism, as opposed to using a flat-sky approximation (which would not be fully adequate here).\footnote{Although one could think that assuming a flat-sky would be a very good approximation for small scales (say, $\ell \gtrsim 50$), the fact that the aberrated power spectrum is given by a convolution of the lensing and un-aberrated power spectra propagates the flat-sky errors to high values of $\ell$  -- see~\cite{hu:2000}.} Note that we are allowed to use a gradient expansion only if we look at angular scales larger than the mean deviation angle. Since in our case the deviation angle is proportional to $\beta$, this means that our treatment is correct for multipoles $\ell<1/\beta\sim 1000$. On smaller scales an exact treatment would be required; this is discussed in the case of weak lensing {\it e.g.} in~\cite{Lewis}. While we do not attempt an exact treatment, the perturbative expression is likely to give a meaningful order of magnitude estimate, as it happens in the case of lensing in~\cite{Lewis}.

Given the temperature anisotropy $\Delta(\hat{{\bf n}})$ and the lensing profile $\Theta(\hat{{\bf n}})$,
we will need their spherical harmonic decompositions, defined respectively
as:
\begin{equation}
    a_{\ell m}\equiv\int d\hat{{\bf n}}~\Delta(\hat{{\bf n}})~Y_{\ell m}^{*}(\hat{{\bf n}})~~~,~~~~~~~~b_{\ell m}\equiv\int d\hat{{\bf n}}~\Theta(\hat{{\bf n}})~Y_{\ell m}^{*}(\hat{{\bf n}})~~~~.\label{almblm}
\end{equation}
This leads to the standard equation for the lensed $a_{\ell m}$:
\begin{equation}
a_{\ell m}=a_{\ell m}^{(P)}+a_{\ell\, m}^{(L)}
\end{equation}
where $a_{\ell m}^{(P)}$ is the unlensed signal and the correction is given by
\begin{equation}\label{eq:almL-gaunt}
    a_{\ell\, m}^{(L)}\simeq\sum_{\ell'\ell''}(-1)^{m}G_{~\ell~\ell'~\ell''}^{-mm0} \frac{\ell'(\ell'+1)-\ell(\ell+1)+\ell''(\ell''+1)}{2}a_{\ell'\, m}^{(P)}b_{\ell''0}\,,
\end{equation}
where we have used the fact that $b_{\ell m}$ is proportional to $\delta_{m0}$ and we kept only the first 2 terms in Eq.~\eqref{templensing} (and thus Eq.~\eqref{eq:almL-gaunt} is only correct at order $\beta$). Indeed, $\Theta(\hat{{\bf n}})$ depends only on the $\theta$ angle if we take the axis of decomposition of the $a_{\ell m}$ in the direction of the velocity $\mathbf{\hat{v}}$. We have also introduced the Gaunt integrals, given in terms of the Wigner 3-j symbols (see \textit{e.g.}~\cite{Bartolo:2004if}) as follows:
\begin{equation}
    G_{\ell_{1}~\ell_{2}~\ell_{3}}^{m_{1}m_{2}m_{3}}\equiv\sqrt{\frac{(2\ell_{1}+1)(2\ell_{2}+1)(2\ell_{3}+1)}{4\pi}}
    \left(
    \begin{array}{ccc}
    \ell_{1} & \ell_{2} & \ell_{3}\\
    0 & 0 & 0
    \end{array}\right)
    \left(
    \begin{array}{ccc}
    \ell_{1} & \ell_{2} & \ell_{3}\\
    m_{1} & m_{2} & m_{3}
    \end{array}\right)~.
\end{equation}
Moreover, at order $\beta$ the aberration term $b_{\ell''0}$ just reduces to the dipole,
\begin{equation}\label{eq:b10}
    b_{10}=2\sqrt{\frac{\pi}{3}}\beta~~,
\end{equation}
leading to:
\begin{equation}
    a_{\ell\, m}^{(L)}\simeq\sum_{\ell'}(-1)^{m}G_{~\ell~\ell'~1}^{-mm0}\frac{\ell'(\ell'+1)-\ell(\ell+1)+2}{2}a_{\ell'\, m}^{(P)}b_{10}\,,\label{aL}
\end{equation}
Because of the properties of the Wigner 3-j symbols, we may write this in the following general form:
\begin{equation}
    a_{\ell\, m}^{(L)}\simeq c_{\ell m}^{-}a_{\ell-1\, m}^{(P)}+c_{\ell m}^{+}a_{\ell+1\, m}^{(P)}\,.
    \label{basic}
\end{equation}
The Gaunt integrals in Eq. (\ref{aL}), simplify considerably leading to the following coefficients:
\begin{eqnarray}
    c_{\ell m}^{+} & = & b_{10}\frac{1}{2}\sqrt{\frac{3}{\pi}}(\ell+2) \sqrt{\frac{(\ell+1)^{2}-m^{2}}{4(\ell+1)^{2}-1}}=\beta(\ell+2)\sqrt{\frac{(\ell+1)^{2}-m^{2}}{4(\ell+1)^{2}-1}}\,,\nonumber \\
    c_{\ell m}^{-} & = & -b_{10}\frac{1}{2}\sqrt{\frac{3}{\pi}}(\ell-1) \sqrt{\frac{\ell^{2}-m^{2}}{4\ell^{2}-1}}=-\beta(\ell-1)\sqrt{\frac{\ell^{2}-m^{2}}{4\ell^{2}-1}}\,.  \label{cABERR}
\end{eqnarray}

As it has been pointed out by \cite{challinor02}, there are also additional terms in this expression due to the Doppler effect. We summarize the calculations performed by \cite{challinor02} in Appendix \ref{app}, which lead to:
\begin{eqnarray}
    c_{\ell m}^{+} & = & \beta(\ell+2 -d)\sqrt{\frac{(\ell+1)^{2}-m^{2}}{4(\ell+1)^{2}-1}}\,,\nonumber \\
    c_{\ell m}^{-} & = & -\beta(\ell-1+ d )\sqrt{\frac{\ell^{2}-m^{2}}{4\ell^{2}-1}}\,,\label{cTOTAL}
\end{eqnarray}
%\begin{eqnarray}
%    c_{\ell m}^{+} & = & -\beta(\ell+2-4d)\sqrt{\frac{(\ell+1)^{2}-m^{2}}{4(\ell+1)^{2}-1}}\nonumber \\
%    c_{\ell m}^{-} & = & \beta(\ell-1+4d)\sqrt{\frac{\ell^{2}-m^{2}}{4\ell^{2}-1}}\label{cTOTAL}
%\end{eqnarray}
where the term due to $d$ contains the pure Doppler effect. Setting the value $d=4$ we would recover the coefficients obtained by~\cite{challinor02} for the multipoles of the intensity integrated over all the frequencies. However as discussed in Appendix~\ref{app}, we are interested experimentally in a slightly different quantity, namely the Thermodynamic Temperature $T$.\footnote{Note in fact that a subtlety of this analysis is that, despite the fact that $I=T^4$, this does not mean that $ \Delta I/I=4\Delta T/T$. In fact this is true only at first order, while at second order we need to include $ \Delta I/I=4\Delta T/T+6(\Delta T/T)^2$. Now, since in the boosted frame $T=T_0(1+\beta \cos(\theta))(1+\Delta T^{(P)}/T)$, then when taking the square in $(\Delta T/T)^2$ we also get a term which goes like $\beta \cos(\theta)(\Delta T^{(P)}/T)$ which represents an additional mixing between neighboring multipoles leading to different coefficients $c_{\ell m}^{+} $ and $c_{\ell m}^{-} $ for the Intensity and the Temperature.} This is derived experimentally from the Intensity at some given frequency $\nu$ via the transformation Eq.~(\ref{T2}) and transforms under a boost as in eq.~(\ref{boost}). In this case the above coefficients take the value $d=1$, in agreement with~\cite{Kosowsky}. From now on all our results will be valid for the Temperature fluctuations.

Note also that the Doppler term is subdominant for large $\ell$; however in an observable quantity such as the two point correlation function, the two coefficients $c_{\ell m}^{+}$ and $c_{\ell m}^{-}$ appear in a combination which leads to a partial cancelation in the leading term, proportional to $\ell$. In the end, as discussed in section \ref{measuringbeta}, it turns out that both effects - aberration and Doppler - contribute in a comparable amount.

The same coefficients can be computed also for $E$ and $B$ polarization fields. We can summarize the general result (see again Appendix \ref{app}) at order $\beta$
\begin{equation}
a_{\ell m}^{X}=a_{\ell m}^{(P_{X})}+c_{\ell m}^{-X}a_{\ell-1m}^{(P_{X})}+c_{\ell m}^{+X}a_{\ell+1m}^{(P_{X})}\label{eq:genform}
\end{equation}
where the index $X$ represents $T$ (temperature), $E$ (electric
component of the polarization) or $B$ (magnetic component of the
polarization). The $T$ coefficients have been displayed above, in
Eq. (\ref{cTOTAL}), while the coefficients for $E$ and $B$ are
given by:
\begin{eqnarray}
c_{\ell m}^{+E,B} & = & \beta(\ell+2-d)\sqrt{\frac{((\ell+1)^{2}-m^{2})(\ell^{2}-4)}{\ell^{2}(4(\ell+1)^{2}-1)}} \,, \label{cTOTAL2}\\
c_{\ell m}^{-E,B} & = & -\beta(\ell-1+d)\sqrt{\frac{(\ell^{2}-m^{2})(\ell^{2}-4)}{\ell^{2}(4\ell^{2}-1)}}\,,
\end{eqnarray}
where, again,  $d$ plays the same role as in the Eqs.~\ref{cTOTAL}.
%\begin{eqnarray}
%c_{\ell m}^{+E,B} & = & -\beta(\ell+2-4d)\sqrt{\frac{((\ell+1)^{2}-m^{2})(\ell^{2}-4)}{\ell^{2}(4(\ell+1)^{2}-1)}}\label{cTOTAL2}\\
%c_{\ell m}^{-E,B} & = & \beta(\ell-1+4d)\sqrt{\frac{(\ell^{2}-m^{2})(\ell^{2}-4)}{\ell^{2}(4\ell^{2}-1)}}
%\end{eqnarray}
For large $\ell$ they reduce to exactly the same coefficients of Eq.~(\ref{cTOTAL}).
Finally it is interesting to note that~\cite{challinor02} also found a (weak) correlation  between the E-mode and the B-mode of the polarization and between the Temperature and the B-mode already at  ${\cal O}(\beta)$.
However, as discussed in Appendix~\ref{app}, when looking at quantities with dimensions of Temperature (instead of Intensities) we find that these correlations exactly disappear: namely the Doppler corrections exactly cancels the Aberration.

%%%%%%%%%%%%%%%%%%%%%%%%%%%%%%%%%%%%%%%%%%%%%%%%%%%%%%%%%%%%%%%%%%%%%%%%%

\section{Measuring $\beta$}

\label{measuringbeta}

As we have seen in the previous section, a mixing between multipoles
$\ell$ and $\ell+1$ is present at order $\beta$, so that the observed
multipoles are corrected as in Eq.~(\ref{eq:genform}). We discuss
now separately the possibility of detecting the aberration signal
in temperature and polarization maps.

\subsection{Temperature}

We can try and construct a non-vanishing two point function which
contains this effect. As discussed in Sec. I, the most straightforward one, the power spectrum, is unaffected by this since it is diagonal in $\ell$ space.
Although there is actually an effect, it is only at order $\beta^{2}$~\cite{challinor02}, which is too small to be observable.
%We discuss it in Appendix ~\ref{appCl}.

The off-diagonal part of the two-point correlation function leads
to a much larger signal, of order $\beta$, instead of $\beta^{2}$.
We define the following basic quantities:
\begin{eqnarray}
F_{\ell m}\equiv a_{\ell\, m}^{*}a_{\ell+1\, m}~~.
\end{eqnarray}
 These quantities are not rotationally invariant, since they depend
on the axis of decomposition of the $a_{\ell m}$'s. We assume here to know already the direction of the velocity of the observer  $\mathbf{\hat{v}}$ of the velocity
and we choose it to be the axis of decomposition $\mathbf{\hat{z}}$. We can in this way detect the modulus $\beta$ of the velocity.
We discuss in section~\ref{sec:dirvel} how to detect the direction of the velocity itself.

We may analyze the two point function above obtaining (at
order $\beta$):
\begin{eqnarray}
F_{\ell m} & = & {a_{\ell\, m}^{(P)}}^{*}a_{\ell+1\, m}^{(P)}+c_{\ell+1m}^{-}{a_{\ell\, m}^{(P)}}^{*}a_{\ell\, m}^{(P)}+c_{\ell m}^{+}{a_{\ell+1\, m}^{(P)*}}a_{\ell+1\, m}^{(P)}\nonumber \\
 & + & c_{\ell m}^{-}{a_{\ell-1\, m}^{(P)}}^{*}a_{\ell+1\, m}^{(P)}+c_{\ell+1m}^{+}{a_{\ell\, m}^{(P)*}}a_{\ell+2\, m}^{(P)}\,.
\end{eqnarray}
Taking now a statistical average we get simply:
%\begin{eqnarray}
%\langle F_{\ell m}\rangle & = & c_{\ell+1m}^{-}\langle{a_{\ell\, m}^{(P)}}^{*}a_{\ell\, m}^{(P)}\rangle+c_{\ell m}^{+}\langle{a_{\ell+1\, m}^{(P)*}}a_{\ell+1\, m}^{(P)}\rangle=\nonumber \\
% & = & c_{\ell+1m}^{-}C_{\ell}+c_{\ell m}^{+}C_{\ell+1}~~, \label{flm}
%\end{eqnarray}
\begin{eqnarray}
\langle F_{\ell m}\rangle \;=\; c_{\ell+1m}^{-}\langle{a_{\ell\, m}^{(P)}}^{*}a_{\ell\, m}^{(P)}\rangle+c_{\ell m}^{+}\langle{a_{\ell+1\, m}^{(P)*}}a_{\ell+1\, m}^{(P)}\rangle \;=\; c_{\ell+1m}^{-}C_{\ell}+c_{\ell m}^{+}C_{\ell+1}~~, \label{flm}
\end{eqnarray}
where $C_{\ell}\equiv\langle a_{\ell\, m}^{*}a_{\ell\, m}\rangle\approx\langle{a_{\ell\, m}^{(P)}}^{*}a_{\ell\, m}^{(P)}\rangle$
is the multipole power spectrum. From Eq.~(\ref{flm}) one sees that for a static observer $\langle F_{\ell m}\rangle$ is equal to zero.\footnote{This is true in linear cosmological perturbation theory. However, non linear effects may induce non zero correlations between different multipoles. It would be therefore interesting to compare
 the correlation patterns emerging in this case with the correlations studied in the present paper.}  Note also that this effect does not have an analogous counterpart in usual weak lensing. In fact, in lensing studies the potentials have vanishing statistical average, so that the first nonzero effect arises only at the quadratic level in the potentials. Here instead we have used the fact that the velocity has a nonzero known value $\langle\beta\rangle=\bar{\beta}\neq0$.

For the following it will be useful to work in terms of real quantities:
\begin{equation}
\frac{1}{2}(F_{\ell m}+F_{\ell-m})\equiv f_{\ell m},\,\qquad\frac{1}{2i}(F_{\ell m}-F_{\ell-m})\equiv g_{\ell m}\,.
\end{equation}
Since $f_{\ell m} = f_{\ell -m}$, for each $\ell$ we have $\ell+1$ independent real numbers $f_{\ell m}$ ($\ell\geq m\geq0$), and $\ell$ independent real numbers $g_{\ell m}$. While we have no prediction for the $g_{\ell m}$ (\textit{i.e.} their mean value is $\langle g_{\ell m}\rangle=0$), we do have one for the $f_{\ell m}$:
%Since $\ell\geq m\geq0$, ($f_{\ell m} = f_{\ell -m}$), for each $\ell$ we have $\ell+1$ independent real numbers, $f_{\ell m}$, and $\ell$ independent real numbers, $g_{\ell m}$.
\begin{eqnarray}
f_{\ell m}^{TH}\;\equiv\;\langle f_{\ell m}\rangle \;=\; \langle c_{\ell+1m}^{-}{a_{\ell\, m}^{\,*}}a_{\ell\, m}+c_{\ell m}^{+}a_{\ell+1\, m}^{*}a_{\ell+1\, m}\rangle \;=\; c_{\ell+1m}^{-}C_{\ell}+c_{\ell m}^{+}C_{\ell+1}~~.\label{fth}
\end{eqnarray}

In the Gaussian hypothesis, the $f_{\ell m}$ have at lowest order a diagonal correlation matrix:
\begin{equation}
\langle f_{\ell m}^{}f_{\ell'm'}\rangle=\langle F_{\ell m}^{}F_{\ell'm'}\rangle
\approx \frac{1}{2} C_{\ell}C_{\ell+1}\delta_{\ell\ell'} \delta_{mm'} (1+\delta_{m0})\,,
\label{eq:fullcorr}\end{equation}
so that the total signal is simply the sum over all the $\ell,m$'s.
Each $f_{\ell m}$ has a cosmic variance given by
$\sigma_{\ell m}^{2}=\langle f_{\ell m}^{2}\rangle-\langle f_{\ell m}\rangle^{2}$
, that is \begin{equation}
\sigma_{\ell m}^{2}=\frac{1}{2} C_{\ell}C_{\ell+1}(1+\delta_{m0})-(f_{\ell m}^{TH})^{2}
\approx \frac{1}{2}C_{\ell}C_{\ell+1}(1+\delta_{m0})\,
\end{equation}
(since $f_{\ell m}^{TH}$ is of order $\beta$). We can now build
a signal-to-noise ratio summing over all $\ell,m$:\footnote{Starting from a sum over $m\ge 0$ we then derive an expression summed over all $m$'s by using the identity $f_{\ell m} = f_{\ell -m}$ and Eq.~(\ref{eq:fullcorr}). The same for Eq.~(\ref{chi2beta}).}
\begin{equation}
\left(\frac{S}{N}\right)^{2}=\sum_{\ell,m}\,\frac{(f_{\ell m}^{TH})^{2}}{C_{\ell}C_{\ell+1}}~~.\label{SN}
\end{equation}
Here, and in the following, we adopt the convention that a sum over
$\ell,m$ is to be performed first over the range $m\in(-\ell,\ell)$
and then over $\ell\in(2,\ell_{max})$, except where otherwise indicated.
If an experiment finds a signal with a $(S/N)$ larger than, say, 3, we can safely
assume that a detection has been made.

\begin{figure}[h]
\begin{centering}
\begin{tabular}{c}
    \includegraphics[width=15cm]{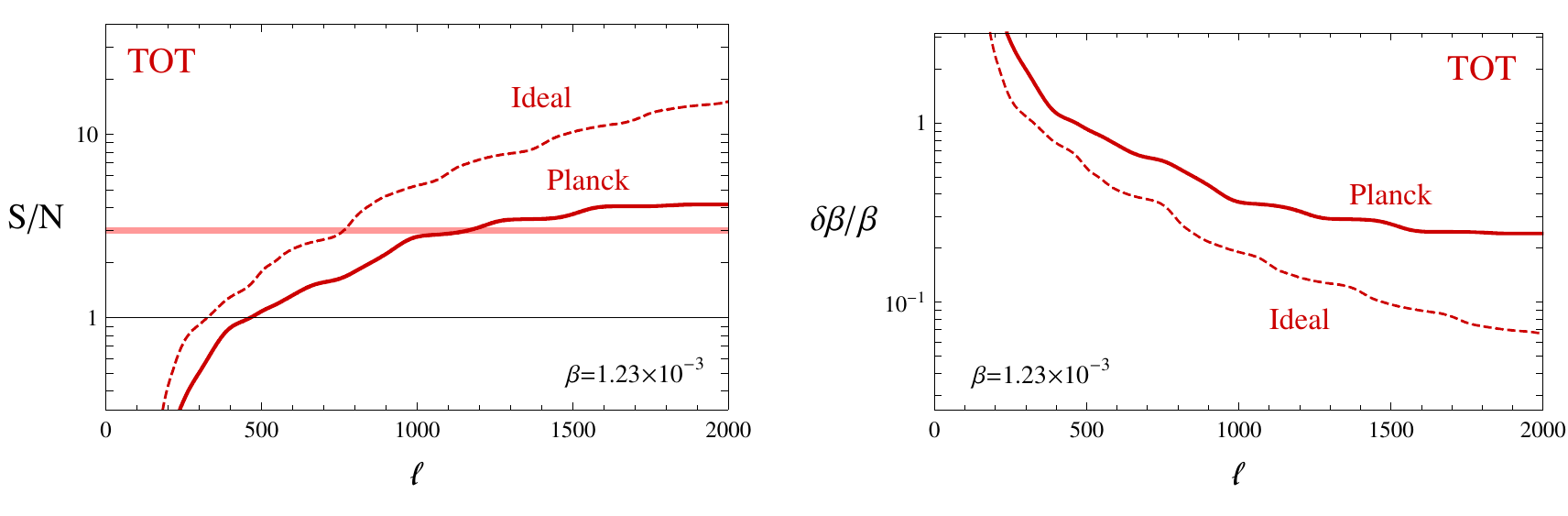}  \tabularnewline
\end{tabular}
\end{centering}
\caption{Signal-to-noise ratio for non diagonal temperature correlation function assuming $\beta=1.23\times10^{-3}$. Left: ideal case with full computation (solid), approximation as described in the text (dot-dashed), lensing contribution only (dotted). Right: ideal case (dotted) and Planck experiment (solid): $f_{sky}=0.85$ and noise as described in the text. The horizontal line represents the detection threshold S/N=3. \label{fig-LocVel-SNnd}}
\end{figure}

For an ideal experiment, this signal-to-noise ratio, including lensing
and Doppler effects simultaneously, is shown by the upper curve in
Fig.~\ref{fig-LocVel-SNnd} (both panels). We have fixed $\beta=1.23\times10^{-3}$.
On the left panel: the dot-dashed curve is obtained by approximating
$C_{\ell}$ with $C_{\ell+1}$ in Eq. (\ref{fth}); the lower dotted
curve shows how the signal-to-noise changes by considering only the
lensing contribution, Eq. (\ref{cABERR}). The right panel shows the
comparison between an ideal and a real experiment. Indeed, the solid
curve includes the noise as foreseen for the Planck experiment, as
well as a sky cut. This is made by replacing the $C_{\ell}$'s in
the denominator with the following quantities ${\mathfrak C}_{\ell}$ (see \textit{e.g.} \cite{lesgourgues}):
\begin{equation}
{\mathfrak C}_{\ell}=\frac{1}{f_{sky}}(C_{\ell}+N_{\ell})\;, \qquad N_{\ell} =\theta_{0}^{2}\sigma_{T}^{2}e^{\ell(\ell+1)\theta_{0}^{2}/(8\log{2})}\,,\label{eq:noise}
\end{equation}
 where we take for Planck~\cite{lesgourgues,Melchiorri}
the sensitivity of the best channel:
\begin{equation}
\sigma_{T}=2\times10^{-6}\;,\qquad\theta_{0}=7'\;,\qquad f_{sky}=0.85\,.
\end{equation}
It turns out that an experiment which goes up to $\ell_{max}\gtrsim 1000$, such as Planck,
has the possibility of observing the effect with a high Signal-to-Noise ratio.
However we should keep in mind that in order to use multipoles $\ell>1/\beta$ in
 a prediction an exact calculation is needed, which goes beyond the gradient expansion used in this paper, as discussed in sect.~\ref{sec:alm}.

The best fit value for $\beta$ in a given set of observed $a^{\rm obs}_{\ell m}$,
decomposed along the axis of the dipole, is given by the minimization
of the $\chi^{2}$:
\begin{eqnarray}
\chi^{2}(\beta)=\sum_{\ell,m}\frac{\left[f_{\ell m}^{\rm obs}-\beta\hat{f}_{\ell m}^{TH}\right]^{2}}{{\mathfrak C}_{\ell}{\mathfrak C}_{\ell+1}}\;,\label{chi2beta}
\end{eqnarray}
 where we neglected terms of order $\beta^{2}$ at denominator and
we have extracted the dependence on $\beta$, writing $f_{\ell m}^{TH}\equiv\beta\hat{f}{}_{\ell m}^{TH}$.
This leads to the estimator
\begin{eqnarray}
\hat{\beta}=\left(\sum_{\ell,m}\frac{f_{\ell m}^{\rm obs}\hat{f}_{\ell m}^{TH}}{{\mathfrak C}_{\ell}{\mathfrak C}_{\ell+1}}\right)\left(\sum_{\ell,m}\frac{(\hat{f}_{\ell m}^{TH})^{2}}{{\mathfrak C}_{\ell}{\mathfrak C}_{\ell+1}}\right)^{-1}\;.
%\label{beta}
\end{eqnarray}
 We can now use Eq. (\ref{eq:fullcorr}) to approximate the variance
of $\hat{\beta}$ as follows:\begin{equation}
\frac{(\delta \beta)^{2}}{\beta^{2}}\equiv \frac{\langle\hat{\beta}^{2}\rangle}{\beta^{2}}\approx\left(\sum_{\ell,m}\frac{\beta^{2}(\hat{f}_{\ell m}^{TH})^{2}}{{\mathfrak C}_{\ell}{\mathfrak C}_{\ell+1}}\right)^{-1}=\left(\frac{N}{S}\right)^{2}\label{eq:relerrbeta} \,.
\end{equation}
%where for consistency we approximated (at the lowest order in $\beta$)
%$\sigma_{\beta}^{2}=\langle\hat{\beta}^{2}\rangle-\beta^{2}\approx\langle\hat{\beta}^{2}\rangle$.

\subsection{Polarization}

We can generalize easily the above consideration to the polarization measurements.
The observable quantities are defined as: \begin{eqnarray}
\langle F_{\ell m}^{XY}\rangle & \equiv & \langle a_{\ell\, m}^{X*}a_{\ell+1\, m}^{Y}\rangle=c_{\ell+1m}^{-Y}\langle{a_{\ell\, m}^{(P_{X})}}^{*}a_{\ell\, m}^{(P_{Y})}\rangle+c_{\ell m}^{+X}\langle{a_{\ell+1\, m}^{(P_{X})}}^{*}a_{\ell+1\, m}^{(P_{Y})}\rangle\nonumber \\
 & = & c_{\ell+1m}^{-Y}C_{\ell}^{XY}+c_{\ell m}^{+X}C_{\ell+1}^{XY} \, ,
 \end{eqnarray}
where $X,Y$ can be either Temperature ($T$) or polarization ($E$ or $B$).

\begin{figure}[t]
\begin{centering}
\begin{tabular}{c}
\includegraphics[width=15cm]{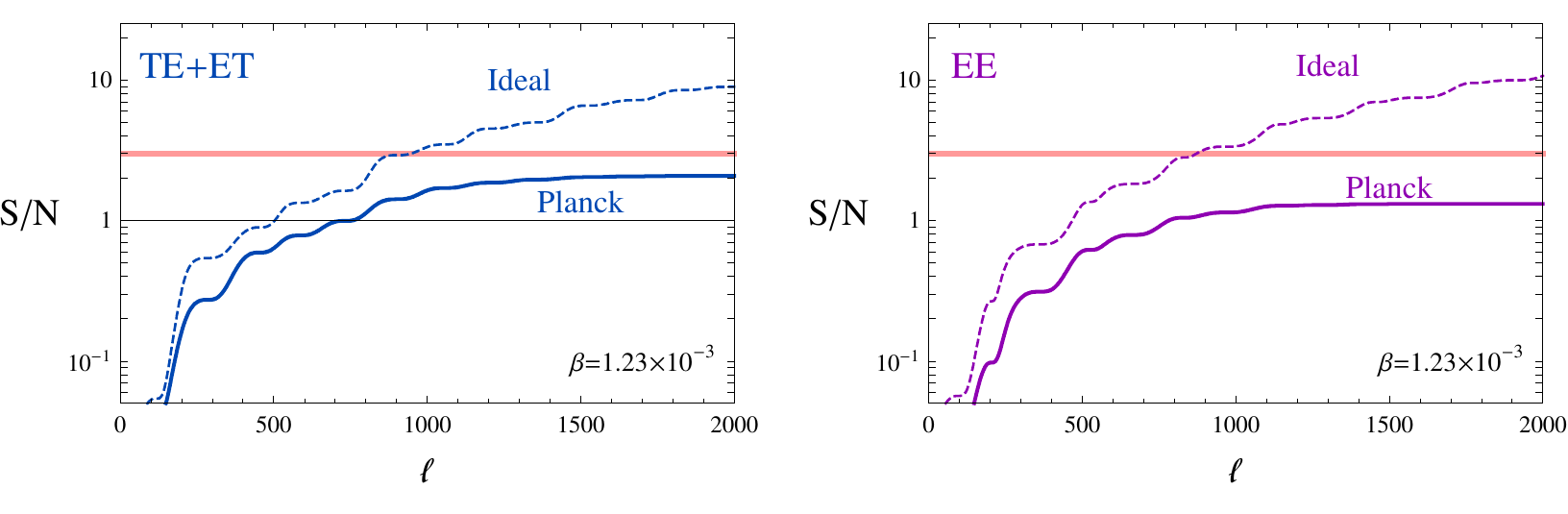} \tabularnewline
\end{tabular}
\par\end{centering}
\caption{Signal-to-noise ratio for non diagonal power spectrum for polarization.
The dotted curves refer to the ideal case (no noise), while the solid
curve gives the result for the Planck experimental setup (see description
in the text).}
\label{figPol}
\end{figure}

 It is useful again to define real quantities: \begin{equation}
\frac{1}{2}(F_{\ell m}^{XY}+F_{\ell-m}^{XY})\equiv f_{\ell m}^{XY},\,\qquad\frac{1}{2i}(F_{\ell m}^{XY}-F_{\ell-m}^{XY})\equiv g_{\ell m}^{XY}\,.\end{equation}
 Again, for each $\ell$ we have $\ell+1$ real numbers $f_{\ell m}$
and $\ell$ real numbers $g_{\ell m}$. Again we have no prediction
for the $g_{\ell m}$ (\textit{i.e.} their mean value is $\langle g_{\ell m}\rangle=0$),
while we have a prediction for the $f_{\ell m}$: \begin{eqnarray}
\langle f_{\ell m}^{XY}\rangle=c_{\ell+1m}^{-Y}C_{\ell}^{XY}+c_{\ell m}^{+X}C_{\ell+1}^{XY}\label{eq:lin_comb}\end{eqnarray}
 In addition to $C_{\ell}^{TT}$ which has been already discussed,
there are three nonzero primordial correlators: $C_{\ell}^{TE}$,
$C_{\ell}^{EE}$ and $C_{\ell}^{BB}$:
\begin{equation}
\begin{aligned}
    \langle a_{\ell_{1}m_{1}}^{T*}a_{\ell_{2}m_{2}}^{E}\rangle & \;=\;  C_{\ell}^{TE}\delta_{\ell_{1}l_{2}}\delta_{m_{1}m_{2}}\,, \\
    \langle a_{\ell_{1}m_{1}}^{E*}a_{\ell_{2}m_{2}}^{E}\rangle & \;=\;  C_{\ell}^{EE}\delta_{\ell_{1}l_{2}}\delta_{m_{1}m_{2}}\,, \\
    \langle a_{\ell_{1}m_{1}}^{B*}a_{\ell_{2}m_{2}}^{B}\rangle & \;=\;  C_{\ell}^{BB}\delta_{\ell_{1}l_{2}}\delta_{m_{1}m_{2}}\;.
\end{aligned}
\end{equation}
 This leads to the following possible observables: $f_{\ell m}^{TT}$,
$f_{\ell m}^{TE}$, $f_{\ell m}^{ET}$, $f_{\ell m}^{EE}$ and $f_{\ell m}^{BB}$.
Their specific shape depends on the form of the primordial correlators.
Note also that: \begin{eqnarray}
\langle f_{\ell m}^{XY}f_{\ell m}^{HK}\rangle=\frac{1}{2}{ C}_{\ell}^{XH}{ C}_{\ell+1}^{YK}\left(1+\delta_{m0}\right)\,.\label{corr}\end{eqnarray}
 We can  then define a signal-to-noise ratio for the channel $XY$
as follows: \begin{equation}
\left(\frac{S}{N}\right)_{XY}^{2}=\sum_{\ell,m}\,\frac{\langle f_{\ell m}^{XY}\rangle^{2}}{{\mathfrak C}_{\ell}^{XX}{\mathfrak C}_{\ell+1}^{YY}}\,.\label{SNpol}\end{equation}
 We plot in Fig. \ref{figPol} the signal-to-noise ratio for the polarization
channels: $ET$, $TE,$ and $EE$. Note also that, since $TE$ and $ET$
are practically identical, we plot only their sum, defined as the
square root of $\left(\frac{S}{N}\right)_{TE}^{2}+\left(\frac{S}{N}\right)_{ET}^{2}$.
%The $BB$ signal is much smaller than the noise and therefore we do not consider it.
%We show the signal-to-noise ratio of the various channels, analogous to Eq. (\ref{SN}), in
In Fig. \ref{figPol}, the dotted curve shows the situation for an
ideal experiment, while the solid one includes, in addition to the
cosmic variance, the noise of the Planck experiment (where we assumed
that the dominant noise correlations are diagonal in $\ell$) and
the fraction of the sky. For $C_{\ell}^{EE}$, the procedure is the same
already described for the temperature, but with $\sigma_{T}$ replaced
by $\sigma_{E}=\sqrt{2}\,\sigma_{T}$. Note that, since the expected
$BB$ signal is suppressed with respect to the experimental noise,
this channel is highly suppressed, leading to a signal-to-noise of
at most $10^{-2}$ and therefore not detectable with Planck.

\begin{figure}[t]
\begin{centering}
\begin{tabular}{c}
\!\includegraphics[width=15cm]{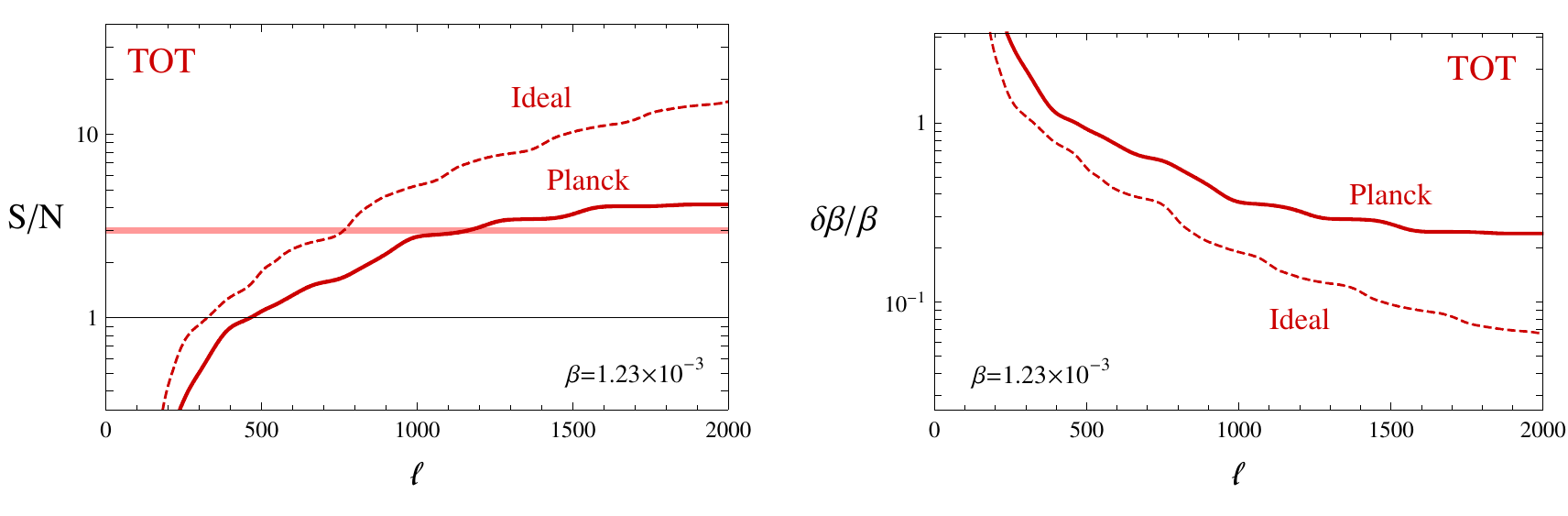} \tabularnewline
\end{tabular}
\par\end{centering}
\caption{Left: Total signal-to-noise ratio for non diagonal power spectrum. Right: relative error $\delta \beta/\beta$.
We fixed $\beta$ at its fiducial value $1.23\times 10^{-3}$.}
\label{figTOT}
\end{figure}

We may finally sum the four off-diagonal temperature and polarization signal-to-noises,
in order to obtain a total one: \begin{equation}
S/N=\sqrt{\sum_{XY}(S/N)_{XY}^{2}}~~~~,~~~~X,Y=T,E\label{eq:totalsn}\end{equation}
 We show the result in the left plot of Fig. \ref{figTOT}. Clearly, this procedure
is not exact, since we are ignoring the correlations between different
components, which could lead to a degradation of the signal. One should
carry out a Fisher matrix analysis, as we actually do in section \ref{Fisher}.
It turns out that, since the correlation matrix is diagonalised by
relatively small rotations (the elements of the correlation matrix
are given in Eq.~(\ref{corr})), the above estimate for the total
signal can be considered quite good, also because the total signal-to-noise
is essentially dominated by the $TT$ contribution (for which the
correlations are small). It turns out that, if $\beta$ is close to its fiducial value, $\beta=1.23\times 10^{-3}$, the Planck experiment should
find a $S/N$ in the range between $5$ and $10$.

Remember that the $S/N$ ratio is linearly proportional to $\beta$. Clearly, there could be a discrepancy between the fiducial value of $\beta$ measured through the dipole, and the value obtained our procedure, namely through the non diagonal correlations of high multipoles. It is then important to understand the precision on $\beta$ provided by our method. In the right plot of Fig.~\ref{figTOT}, we show the relative error for $\beta$, {\it i.e.} $\delta\beta/\beta$, expected assuming that $\beta$ is actually equal to its fiducial value. For other values, just remember that $\delta\beta/\beta$ scales as $\beta^{-1}$. The plot shows that, within our procedure, $\beta$ could be measured by Planck with an error  $\sim 30\%$ (corresponding to a $S/N$ about $4$), whereas an ideal experiment going up to $\ell = 2000$ the error could in principle be reduce to $\sim 7\%$ (corresponding to a $S/N$ about $15$). One should keep in mind though that as discussed in section~\ref{sec:alm}, any estimate going beyond $\ell\sim 1000$ would require an exact treatment of lensing which goes beyond our approach here.

Finally, a $\chi^{2}(\beta)$ can also be constructed for polarization,
in an analoguous way as Eq. (\ref{chi2beta}). The final estimator for
$\beta$ is obviously (again neglecting the correlation among channels)
\begin{eqnarray}
\hat{\beta}=\left(\sum_{X}\sum_{\ell,m}\frac{f_{\ell m}^{X \, {\rm obs}}\langle\hat{f}_{\ell m}^{X}\rangle}{{\mathfrak C}_{\ell}^{X}{\mathfrak C}_{\ell+1}^{X}}\right)\left(\sum_{X}\sum_{\ell,m}\frac{\langle\hat{f}_{\ell m}^{X}\rangle^{2}}{{\mathfrak C}_{\ell}^{X}{\mathfrak C}_{\ell+1}^{X}}\right)^{-1}~~.\label{beta}
\end{eqnarray}
while the final expression in the variance equation (\ref{eq:relerrbeta})
remains the same with $(S/N)$ given in Eq. (\ref{eq:totalsn}).

\section{Measuring the direction of the velocity}

\label{sec:dirvel} The above signal-to-noise ratio applies if we know from the beginning which is the axis $\mathbf{\hat{v}}$ of the velocity and identify it with $\mathbf{\hat{z}}$. However we would like to detect simultaneously the modulus {\it and} direction of the velocity. In general, if we want to work in another frame, with $\mathbf{\hat{z}}'$ oriented along a different axis than $\mathbf{\hat{v}}$, we have to apply a rotation to the $a_{\ell m}^{X}$ coefficients, using the Wigner rotation matrices $D_{m m'}^{\ell} (\phi,\theta,\gamma)$ defined in term of three Euler rotation angles $\phi, \, \theta$ and $\gamma$ (see for instance~\cite{Wignerbook,durrerbook}). We call $\theta,\,\phi$ the angles that identify the direction of $\mathbf{\hat{z}}'$ in the original frame (the one with~$\mathbf{\hat{z}}$). Therefore the parameters in the Wigner matrices can be written simply as
% or equivalently
$D_{m m'}^{\ell}(\phi,\theta,0)$, where $\gamma$, being the angle between the line of nodes and the x-axis, does not affect the direction
of the velocity and can be set to 0. The rotated coefficients are thus~\cite{durrerbook}
\begin{equation}\label{eq:alm-rotated}
\bar{a}_{\ell m}^{X}=\sum_{-\ell\leq m'\leq\ell}D_{m m'}^{\ell}a_{\ell m'}^{X}\,.\end{equation}

In order to build the estimator we define (we drop from now on the bar for the $a_{\ell m}^{X}$ coefficients in the rotated frame)
\begin{align}\label{eq:efdef-1}
    F_{\ell \, m}^{s X Y} \equiv a_{\ell \, m}^{* X}  a_{\ell+1 \, m+s}^{Y}\,.
\end{align}
 where $s \in \{-1, 0 ,1\}$ (although we will restrict only to $s \in \{0 ,1\}$ later). In Appendix~\ref{app:gendir} we find that
\begin{align}
    \left<F_{\ell \, m}^{s X Y} \right>  \,=\, \beta_s \left[c_{\ell+1 \, m+s}^{s\,-\,Y}\, C_\ell^{X Y} + (-1)^s  c_{\ell \, m}^{-s\,+\,X} \,C_{\ell+1}^{X Y} \right]\,,
\end{align}
where $\beta_s \equiv \sqrt{4\pi/3} \,\,Y^*_{1s}$ and the coefficients $c_{\ell \, m}^{s\,\pm,\,X}$ are defined in Appendix~\ref{app:gendir}. A powerful result is that each $\left<F_{\ell \, m}^{s X Y} \right>$  depends only on a single $\beta_s$. This feature naturally implies that the Fisher Matrix can be written in diagonal form, as will be shown below.
%This feature naturally implies that the Fisher Matrix is a diagonal matrix.
%Note that $F_{\ell m}^{\pm\,X Y *}=-F_{\ell -m}^{\mp\,X Y}$.

Since it is convenient to work with real quantities, we define the real and imaginary parts of $F_{\ell \, m}^{s X Y}$ by:
\begin{align}
    f_{\ell \, m}^{s X Y} \,\equiv\, \frac{1}{2} \left[ F_{\ell \, m}^{s X Y}  + \left(F_{\ell \, m}^{s X Y}\right)^* \right]\,, \\
    g_{\ell \, m}^{s X Y} \,\equiv\, \frac{1}{2i} \left[ F_{\ell \, m}^{s X Y}  - \left(F_{\ell \, m}^{s X Y}\right)^* \right]\,.
\end{align}
These quantities in turn satisfy the following (see Appendix~\ref{app:gendir}):
\begin{equation}\label{eq:realimm1-2}
\begin{aligned}
    \left<f_{\ell \, m}^{s X Y} \right>  \,=\, {\textrm{Re}}\left[\beta_s\right] \left[c_{\ell+1 \, m+s}^{s\,-\,Y}\, C_\ell^{X Y} + (-1)^s  c_{\ell \, m}^{-s\,+\,X} \,C_{\ell+1}^{X Y} \right] \,,  \\
    \left<g_{\ell \, m}^{s X Y} \right>  \,=\, {\textrm{Im}}\left[\beta_s\right] \left[c_{\ell+1 \, m+s}^{s\,-\,Y}\, C_\ell^{X Y} + (-1)^s  c_{\ell \, m}^{-s\,+\,X} \,C_{\ell+1}^{X Y} \right] \,,
\end{aligned}
\end{equation}
and
%\begin{align}
%    \left<f_{\ell \, m}^s \, f_{\ell' \, m'}^{s'}\right> - \Big<f_{\ell \, m}^s\Big> \Big<f_{\ell' \, m'}^{s'}\Big> \,&=\, \frac{1}{2} C_\ell \,C_{\ell+1} \,\delta_{\ell \ell'} \delta_{s s'} \left( \delta_{m m'}+\delta_{s 0}\delta_{m 0} \right)\,, \\
%    \left<g_{\ell \, m}^s \, g_{\ell' \, m'}^{s'}\right> - \Big<g_{\ell \, m}^s\Big> \Big<g_{\ell' \, m'}^{s'}\Big> \,&=\, \frac{1}{2} C_\ell \,C_{\ell+1} \,\delta_{\ell \ell'} \delta_{s s'} \left(\delta_{m m'}-\delta_{s 0}\delta_{m 0} \right)\,, \\
%    \left<f_{\ell \, m}^s \, g_{\ell' \, m'}^{s'}\right> - \Big<f_{\ell \, m}^s\Big> \Big<g_{\ell' \, m'}^{s'}\Big> \,&=\, 0\,.
%\end{align}
\begin{equation}\label{eq:covariance-f-g}
\begin{aligned}
    \left<f_{\ell \, m}^{s\,XY} \, f_{\ell' \, m'}^{s'\,HK}\right> - \Big<f_{\ell \, m}^{s\,XY}\Big> \Big<f_{\ell' \, m'}^{s'\,HK}\Big> \,&=\, \frac{1}{2} C_\ell^{XH} \,C_{\ell+1}^{YK} \,\delta_{\ell \ell'} \delta_{s s'} \left( \delta_{m 0} +\delta_{s 0}\delta_{m -m'} \right)\,,  \\
    \left<g_{\ell \, m}^{s\,XY} \, g_{\ell' \, m'}^{s'\,HK}\right> - \Big<g_{\ell \, m}^{s\,XY}\Big> \Big<g_{\ell' \, m'}^{s'\,HK}\Big> \,&=\, \frac{1}{2} C_\ell^{XH}  \,C_{\ell+1}^{YK} \,\delta_{\ell \ell'} \delta_{s s'} \left(\delta_{m 0}-\delta_{s 0}\delta_{m -m'} \right)\,, \\
    \left<f_{\ell \, m}^{s\,XY} \, g_{\ell' \, m'}^{s'\,HK}\right> - \Big<f_{\ell \, m}^{s\,XY}\Big> \Big<g_{\ell' \, m'}^{s'\,HK}\Big> \,&=\, 0\,,
\end{aligned}
\end{equation}
where in the last equation we generalized allowing also for polarization.

Below we forecast the precision on the parameter determination for a given
experiment, such as Planck.

%%%%%%%%%%%%%%%%%%%%%%%%%%%%%%%%%%%%%%%%%%%%%%%%%%%%%%%%%%%%%%%%%%%

\subsection{Fisher matrix analysis}

\label{Fisher}

We address here the question of the error in the determination of $\beta$ and the direction angles of the velocity with a Fisher Matrix analysis.

Through the Cramer-Rao inequality, the Fisher information Matrix encodes
the minimum statistical error one is able to recover on the theoretical
parameters of the model. This is done under the assumption of Gaussian-distributed
signal and noise and that the data sets are independent. If $\mathcal{L}({\mathbf{x}}|{\mathbf{p}})$
is the probability of observing a set of data ${\mathbf{x}}$ given
the true parameters ${\mathbf{p}}$, i.e.~the fiducial model, the
Fisher Matrix is defined as:
\begin{equation}
{\mathbf{F}}_{ij}=-\Big<\frac{\partial^{2}\ln{\mathcal{L}}}{\partial p_{i}p_{j}}\Big>,\label{eq:fish}
\end{equation}
 where the average is taken over the set of data. Assuming that the
likelihood function ${\mathcal{L}}$ is Gaussian and that our observables
are well defined by~\eqref{eq:realimm1-2},
our Fisher Matrix is the sum of two matrices ${\mathbf{F}}_{ij}={\mathcal{F}}_{ij}+{\mathcal{G}}_{ij}$
that read: \begin{eqnarray}
{\mathcal{F}}_{ij}=\sum_{\ell, m, s} \sum_{{\mathbf{X}},{\mathbf{Y}}}\frac{\partial\left<f_{\ell m}^{s\,{\mathbf{X}}}\right>({\mathbf{p}})}{\partial p_{i}}\,S_{f{\mathbf{X}}{\mathbf{Y}}}^{-1}\,\frac{\partial\left<f_{\ell m}^{s\,{\mathbf{Y}}}\right>({\mathbf{p}})}{\partial p_{j}}\,,\\
{\mathcal{G}}_{ij}=\sum_{\ell, m, s} \sum_{{\mathbf{X}},{\mathbf{Y}}}\frac{\partial\left<g_{\ell m}^{s\,{\mathbf{X}}}\right>({\mathbf{p}})}{\partial p_{i}}\,S_{g{\mathbf{X}}{\mathbf{Y}}}^{-1}\,\frac{\partial\left<g_{\ell m}^{s\,{\mathbf{Y}}}\right>({\mathbf{p}})}{\partial p_{j}}\,, \,
\end{eqnarray}
where ${\mathbf{X}}=X_{1}X_{2}=TT,EE,TE,ET$ (and so for ${\mathbf{Y}}$)
denotes temperature, E-channel polarization and temperature-polarization
cross-correlations (neglecting the B-channel polarization set of data),
driven by the $(4\times4)$ - covariance matrices $S_{g,f{\mathbf{X}}{\mathbf{Y}}}$. Notice that in general
$\left<f_{\ell m}^{s\,{\mathbf{TE}}}\right> \neq \left<f_{\ell m}^{s\,{\mathbf{ET}}}\right>$.
%,
%since the linear combination in~(\ref{eq:lin_comb}) is not invariant
%under index exchange (the same holds for the imaginary part). The
%covariance matrices can be made explicit, to wit: %\begin{equation}
%S_{f{\mathbf{X}}{\mathbf{Y}}}=\left(\begin{array}{cc}
%\left<f_{\ell m_{1}m_{2}}^{{\mathbf{TT}}{\mathbf{TT}}}{}^{2}\right> & \left<f_{\ell m_{1}m_{2}}^{{\mathbf{TT}}{\mathbf{EE}}}{}^{2}\right>\\
%\\\left<f_{\ell m_{1}m_{2}}^{{\mathbf{TT}}{\mathbf{EE}}}{}^{2}\right> & \left<f_{\ell m_{1}m_{2}}^{{\mathbf{EE}}{\mathbf{EE}}}{}^{2}\right>\end{array}\right)\qquad S_{g{\mathbf{X}}{\mathbf{Y}}}=\left(\begin{array}{cc}
%\left<g_{\ell m_{1}m_{2}}^{{\mathbf{X}}{\mathbf{X}}}{}^{2}\right> & \left<g_{\ell m_{1}m_{2}}^{{\mathbf{X}}{\mathbf{Y}}}{}^{2}\right>\\
%\\\left<g_{\ell m_{1}m_{2}}^{{\mathbf{X}}{\mathbf{Y}}}{}^{2}\right> & \left<g_{\ell m_{1}m_{2}}^{{\mathbf{Y}}{\mathbf{Y}}}{}^{2}\right>\end{array}\right)
%\end{equation}

\begin{equation}
S_{f}=\left(\begin{array}{cccc}
\Xi_{f}^{{\mathbf{TT}},{\mathbf{TT}}} & \Xi_{f}^{{\mathbf{TT}},{\mathbf{EE}}} & \Xi_{f}^{{\mathbf{TT}},{\mathbf{TE}}} & \Xi_{f}^{{\mathbf{TT}},{\mathbf{ET}}}\\
\\\Xi_{f}^{{\mathbf{EE}},{\mathbf{TT}}} & \Xi_{f}^{{\mathbf{EE}},{\mathbf{EE}}} & \Xi_{f}^{{\mathbf{EE}},{\mathbf{TE}}} & \Xi_{f}^{{\mathbf{EE}},{\mathbf{ET}}}\\
\\\Xi_{f}^{{\mathbf{TE}},{\mathbf{TT}}} & \Xi_{f}^{{\mathbf{TE}},{\mathbf{EE}}} & \Xi_{f}^{{\mathbf{TE}},{\mathbf{TE}}} & \Xi_{f}^{{\mathbf{TE}},{\mathbf{ET}}}\\
\\\Xi_{f}^{{\mathbf{ET}},{\mathbf{TT}}} & \Xi_{f}^{{\mathbf{ET}},{\mathbf{EE}}} & \Xi_{f}^{{\mathbf{EE}},{\mathbf{TE}}} & \Xi_{f}^{{\mathbf{EE}},{\mathbf{ET}}}\end{array}\right);\quad S_{g}=\left(\begin{array}{cccc}
\Xi_{g}^{{\mathbf{TT}},{\mathbf{TT}}} & \Xi_{g}^{{\mathbf{TT}},{\mathbf{EE}}} & \Xi_{g}^{{\mathbf{TT}},{\mathbf{TE}}} & \Xi_{g}^{{\mathbf{TT}},{\mathbf{ET}}}\\
\\\Xi_{g}^{{\mathbf{EE}},{\mathbf{TT}}} & \Xi_{g}^{{\mathbf{EE}},{\mathbf{EE}}} & \Xi_{g}^{{\mathbf{EE}},{\mathbf{TE}}} & \Xi_{g}^{{\mathbf{EE}},{\mathbf{ET}}}\\
\\\Xi_{g}^{{\mathbf{TE}},{\mathbf{TT}}} & \Xi_{g}^{{\mathbf{TE}},{\mathbf{EE}}} & \Xi_{g}^{{\mathbf{TE}},{\mathbf{TE}}} & \Xi_{g}^{{\mathbf{TE}},{\mathbf{ET}}}\\
\\\Xi_{g}^{{\mathbf{ET}},{\mathbf{TT}}} & \Xi_{g}^{{\mathbf{ET}},{\mathbf{EE}}} & \Xi_{g}^{{\mathbf{EE}},{\mathbf{TE}}} & \Xi_{g}^{{\mathbf{EE}},{\mathbf{ET}}}\end{array}\right),
\end{equation}
where $\Xi_{f}^{{\mathbf{X}},{\mathbf{Y}}} = \langle f_{\ell m}^{s\,X_{1}X_{2}} f_{\ell m}^{s\,Y_{1}Y_{2}}\rangle$ and $\Xi_{g}^{{\mathbf{X}},{\mathbf{Y}}} = \langle g_{\ell m}^{s\,X_{1}X_{2}} g_{\ell m}^{s\,Y_{1}Y_{2}}\rangle$
are the covariance matrix elements as written in~\eqref{eq:covariance-f-g}, with noise terms
\begin{equation}
    {\mathfrak C}_{\ell}^{TT}=(C_{\ell}^{TT}+N_{\ell}^{TT})/f_{sky}\,,\qquad
    {\mathfrak C}_{\ell}^{EE}=(C_{\ell}^{EE}+N_{\ell}^{EE})/f_{sky}\,,\qquad
%    {\mathfrak C}_{\ell}^{TE}=(\sqrt{{\mathfrak C}_{\ell}^{TT}{\mathfrak C}_{\ell}^{EE}})/f_{sky}\,,
\end{equation}
for temperature, E-channel
polarization and their cross-correlation, respectively. The overall
uncertainty encloses cosmic variance and experimental noise, as in~(\ref{eq:noise}), including a correction due to the galactic cut.

%${\mathfrak C}_{\ell}^{TE}=(C_{\ell}^{TE})/f_{sky}$ \textbf{(MQ: this last equation must be wrong, as $C_{\ell}^{TE}$ oscillates around zero -- what I used in the plot was ${\mathfrak C}_{\ell}^{TE}=(\sqrt{{\mathfrak C}_{\ell}^{TT}{\mathfrak C}_{\ell}^{EE}})/f_{sky}$ which is always greater than zero. Please confirm this.)}

The inverse of the total Fisher matrix will eventually be the parameter
covariance matrix, whose diagonal elements represents the marginalized
variance errors on ${\mathbf{p}}$.

In this Section it is more convenient to build estimators using cartesian coordinates, in which $\hat{\boldsymbol \beta}$ is decomposed into three components $\{\beta_{x},\beta_{y},\beta_{z}\}$ (see Appendix~\ref{app:gendir} for definitions and detailed calculations). As explained in Appendix~\ref{app:gendir} we shall count independent quantities in the following way: $s \in \{0,+1\}$ with
\begin{equation}\label{eq:sum-ranges}
    m \in \{-\ell,\ell\} \;\mbox{ for }\; s=1; \quad m \in \{0,\ell\} \;\mbox{ for }\; s=0 \quad (s=-1 \;\mbox{ discarded entirely}).
\end{equation}
With this convention the key quantity that enters the Fisher Matrix are
\begin{equation}
\begin{aligned} \label{eq:f-g-lms-derivatives}
    \frac{\partial \left<f_{\ell m}^{s X Y}\right>}{\partial\beta_{r}} &\equiv  \left<f_{\ell m}^{s X Y}\right>_{,r} = \left[- \delta_{rx} \frac{\delta_{s1}}{\sqrt{2}} +  \delta_{rz} \delta_{s0} \right]  h^{s X Y}_{\ell m}, \\
    \frac{\partial \left<g_{\ell m}^{s X Y}\right>}{\partial\beta_{r}} &\equiv  \left<g_{\ell m}^{s X Y}\right>_{,r} = \left[\delta_{ry} \frac{\delta_{s1}}{\sqrt{2}} \right] h^{s X Y}_{\ell m},
\end{aligned}
\end{equation}
where the coefficients $h^{s X Y}_{\ell m}$ are given in~\eqref{eq:f-g-h-lms-pol} of Appendix~\ref{app:gendir}.

A $\chi^2$ can be defined as usual, using eq.\eqref{varianze-pol}:
\begin{equation}
    \chi^2(\beta_r)\,=\,\sum_{X,Y}\sum_{\ell,m,s}\frac{\left|f^{s X Y, \,{\rm OBS}}_{\ell m} - \beta_r \left<f^{s X Y}_{\ell m}\right>_{,r}\right|^2}{{\mathfrak C}_{\ell}^{X Y}{\mathfrak C}_{\ell+1}^{X Y}(1+\delta_{s 0}\delta_{m 0})/2}
    \,+\, \sum_{\ell,m,s}\frac{\left|g^{s X Y, \,{\rm OBS}}_{\ell m} - \beta_r \left<g^{s X Y}_{\ell m}\right>_{,r}\right|^2}{{\mathfrak C}_{\ell}^{X Y}{\mathfrak C}_{\ell+1}^{X Y}/2} ,
\end{equation}
where $f^{s X Y,\,{\rm OBS}}_{\ell m}$ and $g^{s X Y,\,{\rm OBS}}_{\ell m}$ are the observed ones.
%Note that in the second sum the $m=s=0$ term does not diverge and is actually zero (see Appendix~\ref{app:gendir}).
The Fisher Matrix can then be written as
\begin{equation}
\begin{aligned}
    \mathbf{F}_{rr'}&\,=\, \mathcal{F}_{rr} + \mathcal{G}_{rr} \,=\, \sum_{X,Y}\sum_{\ell,m,s}\frac{2}{{\mathfrak C}_{\ell}^{X Y}{\mathfrak C}_{\ell+1}^{X Y}} \left[ \frac{\left<f^{s X Y}_{\ell m}\right>_{\!,r}\left<f^{s X Y}_{\ell m}\right>_{\!,r'}}{1+\delta_{s 0}\delta_{m 0}}
    + \left<g^{s X Y}_{\ell m}\right>_{\!,r}\left<g^{s X Y}_{\ell m}\right>_{\!,r'} \right]
    \\
    &\,=\, \sum_{X,Y}\sum_{\ell}\frac{2}{{\mathfrak C}_{\ell}^{X Y}{\mathfrak C}_{\ell+1}^{X Y}} \left[ \sum_{m=0}^\ell \delta_{rz} \frac{\left(h^{0 X Y}_{\ell m}\right)^2}{1+\delta_{m0}} + \sum_{m=-\ell}^\ell(\delta_{rx} + \delta_{ry}) \frac{\left(h^{1 X Y}_{\ell m}\right)^2}{2} \right] \delta_{r r'}\,,
\end{aligned}
\end{equation}
which again is a diagonal matrix. Now, just as we did in deriving~\eqref{SN} we shall extend the first sum also to negative $m$ values, which in practice incurs a factor of 2 and the disappearance of the $\delta_{m0}$ term. We thus have
\begin{equation}\label{eq:fisher-cartesian}
    \mathbf{F}_{rr'}\,=\, \sum_{X,Y}\sum_{\ell}\sum_{m=-\ell}^\ell\frac{1}{{\mathfrak C}_{\ell}^{X Y}{\mathfrak C}_{\ell+1}^{X Y}} \left[ \delta_{rz} \left(h^{0 X Y}_{\ell m}\right)^2 + (\delta_{rx} + \delta_{ry}) \left(h^{1 X Y}_{\ell m}\right)^2 \right] \delta_{r r'}\,.
\end{equation}
Finally, from~\eqref{eq:f-g-h-lms-pol} one can check that:
\begin{equation}\label{eq:Fii-equality}
    \sum_{m=-\ell}^\ell \left(h^{1 X Y}_{\ell m}\right)^2 \;=\; \sum_{m=-\ell}^\ell \left(h^{0 X Y}_{\ell m}\right)^2\,.
\end{equation}
This implies that $\,\mathbf{F}_{xx} = \mathbf{F}_{yy} = \mathbf{F}_{zz}$. In other words, the estimates of the errors along each direction are the same and we only need to obtain one single standard deviation. This result could have been anticipated on physical grounds, as no direction is preferred in this analysis.

Consequently, we can also define the estimator for the three cartesian components
\begin{align}
    \bar{\beta}_{r}&\,=\,\sum_{r'}( \mathbf{F}^{-1})_{rr'} V_{r'} \,=\,\frac{V_{r}}{ \mathbf{F}_{rr}}\,,\\
    V_{r}&\,\equiv\,\sum_{X,Y}\sum_{\ell,  m, s} \frac{2}{{\mathfrak C}_{\ell}^{X Y}{\mathfrak C}_{\ell+1}^{X Y}} \left[\frac{ f^{s X Y,\,{\rm OBS}}_{\ell m} \left<f^{s X Y}_{\ell m}\right>_{\!,r}}{1+\delta_{s 0}\delta_{m 0}}
    + g^{s X Y,\,{\rm OBS}}_{\ell m} \left<g^{s X Y}_{\ell m}\right>_{\!,r} \right] \nonumber\\
    &\,=\, \sum_{X,Y}\,\sum_{\ell}\sum_{m=-\ell}^\ell\frac{1}{{\mathfrak C}_{\ell}^{X Y}{\mathfrak C}_{\ell+1}^{X Y}} \left[ \sqrt{2}  h^{1 X Y}_{\ell m}  \left(- f^{1 X Y,\,{\rm OBS}}_{\ell m} \delta_{rx}  +  g^{1 X Y,\,{\rm OBS}}_{\ell m} \delta_{ry} \right) +  h^{0 X Y}_{\ell m} f^{0 X Y,\,{\rm OBS}}_{\ell m} \,\delta_{rz} \right],
\end{align}
where in the last line we used~\eqref{eq:f-g-lms-derivatives} and the fact that $\,h^{0 X Y}_{\ell m} = h^{0 X Y}_{\ell -m}\,$ to extend the $s=0$ sum to negative values of $m$. The estimator of the magnitude of the velocity is then exactly
\begin{equation}
    \bar{\beta}=\sqrt{\sum_{a}\bar{\beta_{a}}^{2}}
\end{equation}
and we recover the previous result if we align the z-axis with the correct direction. In conclusion, in cartesian coordinates the Fisher Matrix is diagonal, the predicted limits are independent on the fiducial magnitude and direction of $\boldsymbol{\beta}$ and the statistical errors in parameter space are the same for each direction ($\sigma_{x}=\sigma_{y}=\sigma_{z}=\sigma$). From this analysis the angles $\theta$ and $\phi$ can easily be reconstructed from cartesian components.

\begin{figure}[t]
    \begin{centering}
    \begin{tabular}{c}
    \includegraphics[width=7.2cm]{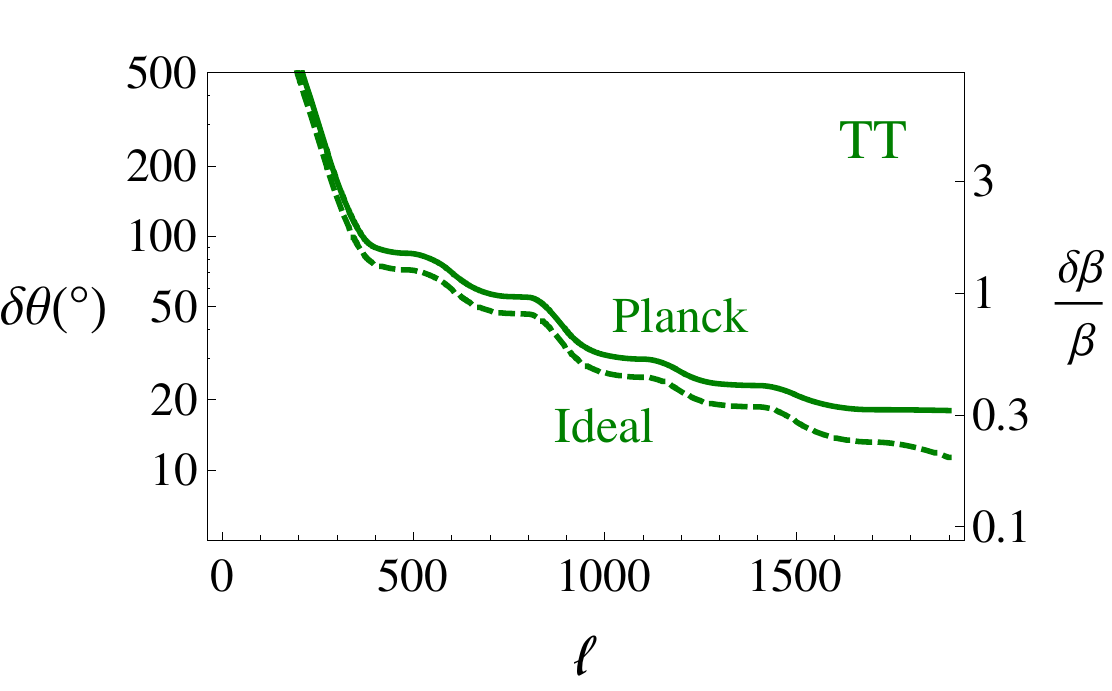} \quad
    \includegraphics[width=7.2cm]{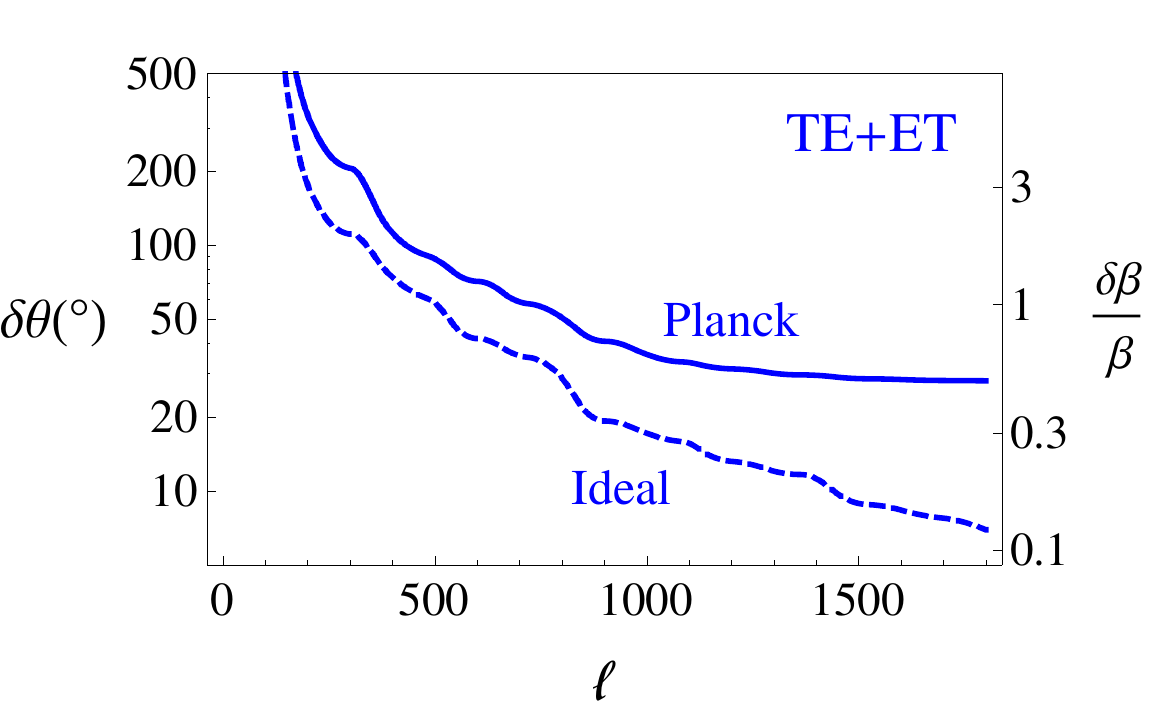}\tabularnewline
    \includegraphics[width=7.2cm]{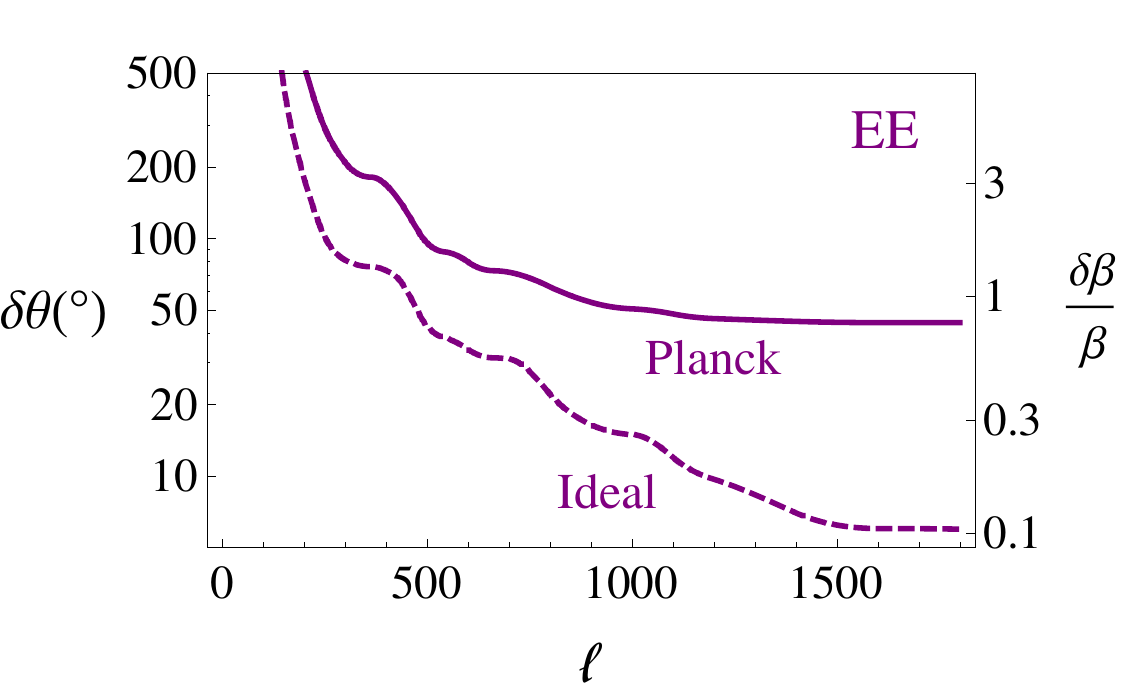} \quad
    \includegraphics[width=7.2cm]{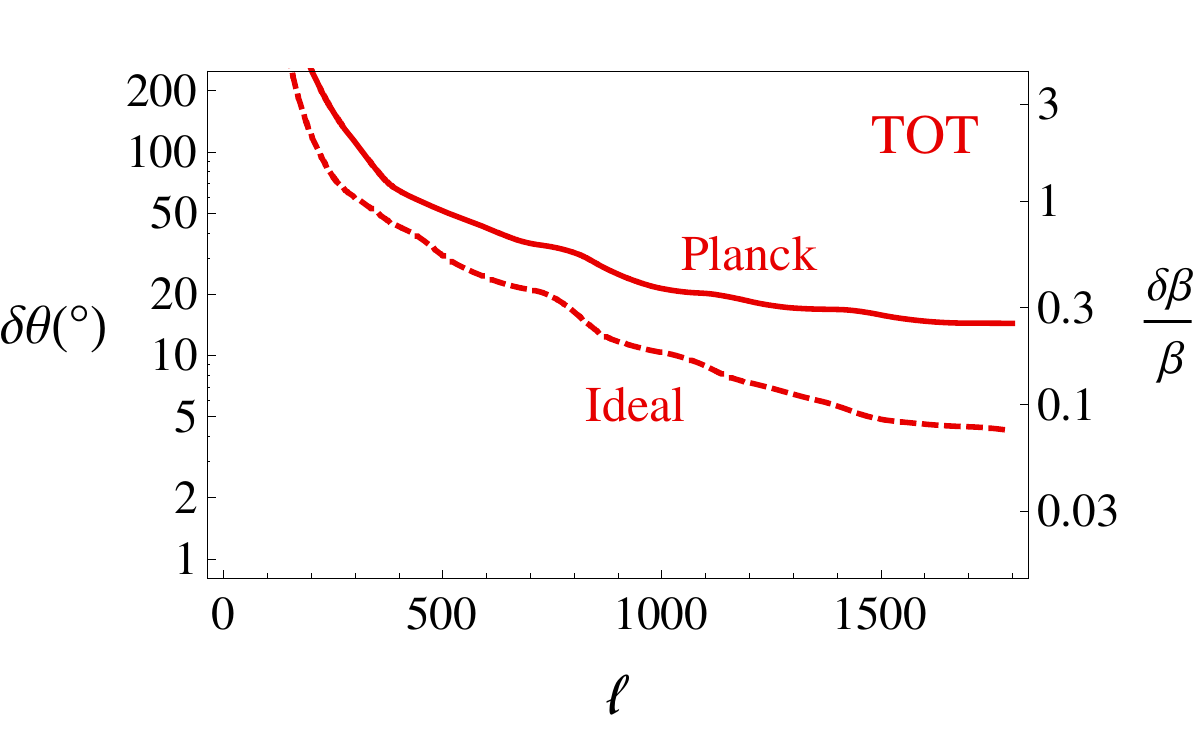}
    \end{tabular}
    \par\end{centering}
    \caption{ Forecast of the error $\delta\theta = \delta\beta/\beta$ (plotted in degrees) in the determination of the direction as a function of the multipole $\ell$ using the TT, TE+ET and EE data, as well as the final error obtained (adding in quadrature) all of these  both for an ideal experiment (dotted line) and for Planck (solid line). No assumption is needed on the fiducial value of $\beta$. Compare with Figure~\ref{figTOT}.}
    \label{TT-TE-EE-thetafig}
\end{figure}

Hence, returning to the case where the fiducial velocity is aligned with the z-axis, we can relate for small angles the error on the absolute value of the velocity with the  error in the magnitude of the direction:
\begin{equation}
\delta\theta=\frac{\sigma}{\beta}=\frac{\delta\beta}{\beta}.
\end{equation}
The final statistical accuracy on $\theta$ and $\beta$ as a function of $\ell_{max}$ is plotted in fig.~\ref{TT-TE-EE-thetafig} for the TT channel and for the polarization channels, TE+ET and EE. The errors in the different channels may be added in quadrature, as a first approximation, as in eq.~\eqref{eq:totalsn}. As it can be seen from the figure, the direction of the velocity can be measured with a precision of about $20^\circ$ for Planck using multipoles up to $\ell\sim 1000$, in good part due to the TT-channel. An ideal experiment with no instrumental noise on the polarization channels could instead go down to a precision of about $4^{\circ}$ if it goes up to $\ell = 2000$, although we remind the reader again that any estimate going beyond $\ell\sim 1000$ would require an exact treatment of lensing beyond first order.

%Missing: Error ellipse plots on the parameters.
%\section{Higher perturbative orders}

%As noted by~\cite{} the leading order correction in~\ref{} is not just of order $\beta$, but of of order $\beta \ell$. This comes from
%We note, however, that the leading order effect in aberration on each $\ell$ is of order $(\beta \ell)^n$ at the $n-{\rm th}$ perturbative level. So, when using $\ell\gtrsim 1000$ extreme care must be taken when interpreting observations. What is needed, in fact, is to compute the aberration to all orders in $\beta$, which we leave for future work.
%This would couple all $\ell's$ together and may lead to severe distortions of the CMB. It is possible to show, for instance, that a model with a sharp feature in the power spectrum, would lead to a ${\cal O}(1)$ distortion on {\it each} $\ell$.

%%%%%%%%%%%%%%%%%%%%%%%%%%%%%%%%%%%%%%%%%%%%%%%%%%%%%%

\section{Conclusions}

In this paper we have shown that a measurement of our peculiar velocity
can be achieved using the information contained in the Cosmic Microwave Background at high multipoles $\ell$, since the velocity of the observer distorts the primordial signal through aberration and Doppler effects.
We have proposed to measure this distortion as a non-zero signal in a two-point correlation function $a_{\ell m_1} a_{\ell+1 m_2}$, which couples neighbouring multipoles $\ell$'s. We have found that the correlation function diagonal in $m$ space can be used to determine the magnitude of the velocity $\beta$, while the off-diagonal one can be used to measure the direction of the velocity and we have forecasted the error on the determination of such quantities using the Temperature and Polarization channels both for an ideal experiment and for the  Planck satellite.

Going up to $\ell\sim1000$, the error on the amplitude of the velocity $\beta$ which can be obtained is of about $30\%-7\%$ (corresponding to a $S/N$ about $4-15$) for Planck and for an ideal experiment, respectively.
The direction of the velocity can also be measured with a precision of about $20^\circ-4^\circ$ respectively for the two cases.

The method employed in the present paper relies on a gradient expansion, which is valid for multipoles $\ell<1/\beta\sim 1000$, while it would require a more refined treatment which does not rely on a gradient expansion for $\ell>1000$. The precision which can be obtained going up to $\ell=2000$ can be estimated with our method and it leads to an improvement of about a factor of 2 in the ideal case, and less than 2 for Planck.

Our proposal constitutes an important independent measurement of our peculiar velocity, which can be compared with the usual measurement, through the Doppler effect on the dipole $\ell=1$. This can be used to distinguish  peculiar velocity from an intrinsic CMB
dipole, provided the latter does not induce at the same time a dipolar
lensing with the same structure in the off-diagonal correlation function.

%Discuss other sources of anisotropy
%This is obtained looking at two-point correlation functions which
%couple two different neighbouring $\ell$'s. %We note, however, that the leading order effect in aberration on each $\ell$ is of order $(\beta \ell)^n$ at the $n-{\rm th}$ perturbative level. So, when using $\ell\gtrsim 1000$ extreme care must be taken when interpreting observations. What is needed, in fact, is to compute the aberration to all orders in $\beta$, which we leave for future work.
%This would couple all $\ell's$ together and may lead to severe distortions of the CMB. It is possible to show, for instance, that a model with a sharp feature in the power spectrum, would lead to a ${\cal O}(1)$ distortion on {\it each} $\ell$.

%%Discuss models with ``jumps'' in the primordial power spectrum.

%%%%%%%%%%%%%%%%%%%%%%%%%%%%%%%%%%%%%%%%%%%%%%%%%%%%

\section*{Acknowledgments}

We would like to thank the anonymous referee for the many useful and detailed suggestions which helped improving the paper, especially on the estimators in cartesian coordinates of Appendix~\ref{app:gendir}. We also thank Paolo Cabella, Michael Doran, Hans Kristian Eriksen, Juan Garcia-Bellido,  Paolo Natoli, Goto Hajime, Cinzia di Porto and Ignacy Sawicki for useful conversations and comments.
%We would like to thank the anonymous referee
%Part of the numerical simulations were performed on the MUST cluster at LAPP (CNRS \& Universit\'e de Savoie).

\section*{Note Added}

When this paper was concluded, we became aware that A. Kosowsky and
T. Kahniashvili in~\cite{Kosowsky} independently obtained results, which are consistent with ours, on the possibility of measuring the magnitude of the velocity $\beta$, analyzing the Temperature coefficients.
In the revised version of the present paper we have decided also to analyze the correlation coefficients for the Temperature fluctuations, as in~\cite{Kosowsky}, instead of the Intensity, as in~\cite{challinor02}. In fact a subtlety of our analysis is that the multipoles for the total integrated Intensity and for the Temperature do not just differ for an overall constant but they have a difference dependence on $\beta$, due to the different Doppler effect.
We have updated all the figures in the present version of the paper, using  Temperature fluctuations, as stressed in sect.~\ref{sec:alm} and Appendix~\ref{app}, and the results now exactly coincide  with~\cite{Kosowsky}.
% for theTT correlations, assuming that the direction is already known.

Then, while~\cite{Kosowsky} has analyzed only the detection of the magnitude of the velocity using the TT correlations, we have also addressed  here the issue of the detection of the direction of the velocity and we have used also the polarization channels.

%%%%%%%%%%%%%%%%%%%%%%%%%%%%%%%%%%%%%%%%%%%%%%%%%%%%%%%%%%%%%%%%%%%%%%%%%%%%%%%%

\appendix

\section{Corrections to $a_{\ell m}^{X}$ due to Doppler and aberration}

\label{app}

We now calculate the modifications induced by the aberration and the
Doppler effect on the quantities $a_{\ell m}^{T}$, $a_{\ell m}^{E}$,
$a_{\ell m}^{B}$, respectively the coefficients of the expansion
in spherical harmonics of the temperature and of the components of the
polarization tensor. To separate the Doppler contribution from the
one associated to the aberration, we introduce a parameter $d$, such
that $d=0$ indicates that the Doppler effect has been neglected and
$d=1$, instead, corresponds to the case in which the interplay between
Doppler and aberration is fully taken in to account.

\subsection{Temperature}

Under a frame transformation the CMB brightness transforms according to
\begin{equation}
I^{\prime}(\nu^{\prime},\hat{\n}^{\prime}) =\left(\frac{\nu^{\prime}}{\nu}\right)^{3d}I(\nu,\hat{\n})\,.\label{I}
\end{equation}
Now, in current CMB experiments, brightness measurements (including those related to polarization) are conventionally translated into an equivalent value for the thermodynamic temperature $T$ through the relation (see for instance~\cite{lewis02,bennett03,partridge})
\begin{equation}
  T = \frac{I(\nu)}{\nu^2}\frac{e^x-1}{x} \equiv \frac{I(\nu)}{\nu^2} f(x) \,,
\label{T}
\end{equation}
where the function $f(x)$ has been defined in the last step and $x$ is a frame invariant quantity given by $x= \nu/T$. For small frequencies, {\it i.e.} $x \to 0$, the function $f(x) \to 1$ and Eq.~(\ref{T}) provides the antenna temperature which by definition is equal to the ratio $I(\nu)/\nu^2$. In general, by solving Eq.~(\ref{T}) one finds
\begin{equation}
T = \frac{\nu}{\log\left(1+\frac{\nu^3}{I(\nu)}\right)} \,,
\label{T2}
\end{equation}
which is the temperature actually measured in a CMB experiment. Since the ratio $\nu^3/I(\nu)$ is frame invariant (see Eq.~(\ref{I}) with $d=1$), from Eq.~(\ref{T2}) one can easily read the transformation properties of the temperature under a Lorentz boost, namely
\begin{equation}
T^{\prime}(\hat{\n}^{\prime}) = \left(\frac{\nu^{\prime}}{\nu}\right)^{d} T(\hat{\n}) \,,
\label{boost}
\end{equation}
where again the parameter $d$ has been introduced to separate the Doppler contribution from the one due to aberration.

We then start our calculation from the following observation \begin{eqnarray}
T^{\prime}(\hat{\n}^{\prime}) & = &
\sum_{\ell^{\prime}m^{\prime}}a_{\ell^{\prime}m^{\prime}}^{T\,\prime}Y_{\ell^{\prime}m^{\prime}}(\hat{\n}^{\prime})\nonumber \\
 & = & \left[\gamma(1+\hat{\n}\cdot{\boldsymbol \beta})\right]^{d}\;\sum_{\ell
m}a_{\ell m}^{T}Y_{\ell m}(\hat{\n})\nonumber \\
 & = & \sum_{\ell m} \sum_{\ell^{\prime}m^{\prime}}\Bigg\{\int
d\underline{\hat{\n}}\,a_{\ell m}^{T}
\left[\gamma(1+\hat{\n}(\underline{\hat{\n}})\cdot{\boldsymbol
\beta})\right]^{d}Y_{\ell^{\prime}m^{\prime}}^{*}(\underline{\hat{\n}})Y_{\ell
m}(\hat{\n}(\underline{\hat{\n}}))\Bigg\}Y_{\ell^{\prime}m^{\prime}}(\hat{\n}^{
\prime})\,,\end{eqnarray}
 where we expanded the function of $\hat{\n}^{\prime}$,
$\left[\gamma(1+\hat{\n}(\hat{\n}^{\prime})\cdot{\boldsymbol \beta})\right]^{d}Y_{\ell
m}(\hat{\n}(\hat{\n}^{\prime}))$,
in the basis $Y_{\ell^{\prime}m^{\prime}}(\hat{\n}^{\prime})$.

This expression leads to the identity
\begin{eqnarray}
  a_{\ell^{\prime}m^{\prime}}^{T\,\prime}& = & \sum_{\ell m} \int
d\underline{\hat{\n}}\,a_{\ell m}^{T}
\left[\gamma(1+\hat{\n}(\underline{\hat{\n}})\cdot{\boldsymbol
\beta})\right]^{d}Y_{\ell^{\prime}m^{\prime}}^{*}(\underline{\hat{\n}})Y_{\ell
m}(\hat{\n}(\underline{\hat{\n}}))\nonumber \\
 & = & \sum_{\ell m} \int d\underline{\hat{\n}}\int
d\hat{\n}^{\prime}\,\delta(\hat{\n}^{\prime}-\underline{\hat{\n}})a_{\ell
m}^{T} \left[\gamma(1+\hat{\n}(\underline{\hat{\n}})\cdot{\boldsymbol
\beta})\right]^{d}Y_{\ell^{\prime}m^{\prime}}^{*}(\underline{\hat{\n}})Y_{\ell
m}(\hat{\n}(\underline{\hat{\n}}))\nonumber \\
 & = & \sum_{\ell m} \int d\hat{\n}^{\prime}\,a_{\ell m}^{T}
\left[\gamma(1+\hat{\n}(\hat{\n}^{\prime})\cdot{\boldsymbol
\beta})\right]^{d}Y_{\ell^{\prime}m^{\prime}}^{*}(\hat{\n}^{\prime})Y_{\ell
m}(\hat{\n}(\hat{\n}^{\prime}))\nonumber \\
 & = & \sum_{\ell m} \int d\hat{\n}\,a_{\ell m}^{T}
\left[\gamma(1+\hat{\n}\cdot{\boldsymbol
\beta})\right]^{d-2}Y_{\ell^{\prime}m^{\prime}}^{*}(\hat{\n}^{\prime})Y_{\ell
m}(\hat{\n})\nonumber \\
 & = & \sum_{\ell} \int d\hat{\n}\,a_{\ell m'}^{T}
\left[\gamma(1+\hat{\n}\cdot{\boldsymbol
\beta})\right]^{d-2}Y_{\ell^{\prime}m^{\prime}}^{*}(\hat{\n}^{\prime})Y_{\ell
m^{\prime}}(\hat{\n})\,,\label{eq:almI}
\end{eqnarray}
where we used the relation $\,d\hat{\n}=\left[\gamma(1+\hat{\n}\cdot{\boldsymbol
\beta})\right]^{2}d\hat{\n}^{\prime}\,$ and in the last line we assumed that the relative velocity ${\boldsymbol \beta}$ points in the direction along which we expanded in spherical harmonics.
In this way a Kronecker delta removes the summation over $m$.

Since we already know that the aberration is a correction which depends
from $\beta$, we expand Eq.(\ref{eq:almI}) in powers of $\beta$
and keep only the leading terms in this expansion. We therefore need
the following Taylor series ($\mu\equiv\hat{\n}\cdot{\boldsymbol \beta}/\beta$
and $\mu^{\prime}\equiv\hat{\n}^{\prime}\cdot{\boldsymbol \beta}/\beta$) \begin{eqnarray}
\left[\gamma(1+\hat{\n}\cdot{\boldsymbol \beta})\right]^{d-2} & = &
\left[(1-\beta^{2})^{-\frac{1}{2}}(1+\mu\beta)\right]^{d-2}
\simeq1+(d-2)\mu\beta+\mathcal{O}(\beta^{2})\nonumber \\
Y_{\ell m}(\hat{\n}^{\prime}) & = & Y_{\ell
m}(\hat{\n}^{\prime})|_{\beta=0}+\frac{Y_{\ell
m}}{\partial\beta}|_{\beta=0}\,\beta+\mathcal{O}(\beta^{2})\nonumber \\
 & = & Y_{\ell m}(\hat{\n}^{\prime})|_{\beta=0}+\left(\frac{\partial Y_{\ell
m}}{\partial\mu^{\prime}}\frac{\partial\mu^{\prime}}{\partial\beta}\right)|_{
\beta=0}\,\beta+\mathcal{O}(\beta^{2})\nonumber \\
 & = & Y_{\ell m}(\hat{\n})+\frac{\partial Y_{\ell
m}}{\partial\mu}(1-\mu^{2})\beta+\mathcal{O}(\beta^{2})\,.
\end{eqnarray}
 By means of the following property of the spherical harmonics \begin{equation}
(\mu^{2}-1)\frac{\partial}{\partial\mu}Y_{\ell m}=\ell
H_{(\ell+1)m}Y_{(\ell+1)m}-(\ell+1)H_{\ell m}Y_{(\ell-1)m}\,,\end{equation}
 where \begin{equation}
H_{\ell m}=\sqrt{\frac{(\ell^{2}-m^{2})}{(4\ell^{2}-1)}}\,,\end{equation}
 we can now write $Y_{\ell m}(\hat{\n}^{\prime})$ as follows \begin{equation}
Y_{\ell m}(\hat{\n}^{\prime})=Y_{\ell m}(\hat{\n})-\beta\Bigg\{\ell
H_{(\ell+1)m}Y_{(\ell+1)m}(\hat{\n})-(\ell+1)H_{\ell
m}Y_{(\ell-1)m}(\hat{\n})\Bigg\}+\mathcal{O}(\beta^{2})\,.\end{equation}

Replacing the previous power series expansions in Eq.(\ref{eq:almI})
one finds (we renamed the indices $\ell$ and $\ell^{\prime}$)
\begin{eqnarray}
a_{\ell m}^{T\,\prime}(\nu^{\prime}) & \simeq & \sum_{\ell^{\prime}}\int d\hat{\n}\,\left[1+(d-2)\mu\beta\right]a_{\ell^{\prime}m}^{T}
Y_{\ell^{\prime}m}(\hat{\n})\nonumber \\
 & \times & \Big[Y_{\ell m}^{*}(\hat{\n})-\beta\Big(\ell H_{(\ell+1)m}Y_{(\ell+1)m}^{*}(\hat{\n})-(\ell+1)H_{\ell m}Y_{(\ell-1)m}^{*}(\hat{\n})\Big)\Big]\nonumber \\
 & \simeq & \sum_{\ell^{\prime}}\int d\hat{\n}\,\Big\{\Big[1+\Big(d-2\Big)\mu\beta\Big]a_{\ell^{\prime}m}^{T}Y_{\ell^{\prime}m}(\hat{\n})Y_{\ell m}^{*}(\hat{\n})\nonumber \\
 & - & \beta a_{\ell^{\prime}m}^{T}(\nu^{\prime})Y_{\ell^{\prime}m}(\hat{\n})\Big[\ell H_{(\ell+1)m}Y_{(\ell+1)m}^{*}(\hat{\n})-(\ell+1)H_{\ell m}Y_{(\ell-1)m}^{*}(\hat{\n})\Big]\Big\}\,.
\end{eqnarray}
The integral over $\dd\hat{\n}$ can be done by using the following
property of the spherical harmonics
\begin{eqnarray}
\int d\hat{\n}\,\mu Y_{\ell^{\prime}m}(\hat{\n})Y_{\ell m}^{*}(\hat{\n}) & = & \int d\hat{\n}\Big\{C_{(\ell^{\prime}+1)m}Y_{(\ell^{\prime}+1)m}(\hat{\n})Y_{\ell m}^{*}(\hat{\n})+C_{\ell^{\prime}m}Y_{(\ell^{\prime}-1)m}(\hat{\n})Y_{\ell m}^{*}(\hat{\n})\Big\}\nonumber \\
 & = & H_{\ell m}\delta_{(\ell^{\prime}+1)l}+H_{(\ell+1)m}\delta_{(\ell^{\prime}-1)l}\,.
 \end{eqnarray}

This leads to the expression \begin{eqnarray}
a_{\ell m}^{T\,\prime}=a_{\ell m}^{T} & - & \beta H_{(\ell+1)m}\Big[(\ell-d+2)\Big]a_{(\ell+1)m}^{T}(\nu^{\prime})\nonumber \\
& - & \beta H_{\ell m}\Big[-(\ell+d-1)\Big]a_{(\ell-1)m}^{T}+\mathcal{O}(\beta^{2}) \nonumber \\
& \equiv & a_{\ell m}^{T}+\left(c_{\ell m}^{-T}a_{\ell-1\, m}^{T}+c_{\ell m}^{+T}a_{\ell+1\, m}^{T}\right) \,.\label{anu}\end{eqnarray}

\subsection{Polarization}

Analogously to the case of the brightness, also the CMB polarization measurements are conventionally translated into units of temperature. As a consequence, it is convenient to introduce a polarization tensor $P^{ab}(\hat{\n})$ with components related to the Stokes parameters and transforming under a Lorentz boost according to Eq.~(\ref{boost}). Its spherical harmonics expansion reads as follows~\cite{challinor02}
\begin{equation}
P^{ab}(\hat{\n}) = \frac{1}{2}\sum_{\ell m}\left(\begin{array}{cc}
(a_{\ell m}^{E}-ia_{\ell m}^{B}){}_{-2}Y_{\ell m}(\hat{\n}) & 0\\
0 & (a_{\ell m}^{E}+ia_{\ell m}^{B}){}_{+2}Y_{\ell m}(\hat{\n})\end{array}\,,\label{Pab}\right)
\end{equation}
 were ${}_{\pm2}Y_{\ell m}$ are the spin-weighted spherical harmonics \cite{kam}.

Expanding in the basis ${}_{\pm2}Y_{\ell m}$ similarly to what we did for the temperature, one finds \begin{equation}
\left[a_{\ell^{\prime} m^{\prime}}^{E\prime}\pm ia_{\ell^{\prime} m^{\prime}}^{B\prime}\right]=\sum_{\ell}\int d\hat{\n}\,\left[a_{\ell m^{\prime}}^{E}\pm ia_{\ell m^{\prime}}^{B}\right] \left[\gamma(1+\hat{\n}\cdot{\boldsymbol \beta})\right]^{d-2}{}_{\pm2}Y_{\ell^{\prime}m^{\prime}}^{*}(\mathbf{\hat{n}}^{\prime}){}_{\pm2}Y_{\ell m^{\prime}}(\hat{\n})\,.\label{eq:alm}
\end{equation}
Then, by means of the following properties of the spherical harmonics
\cite{challinor02}
\begin{eqnarray}
 &  & (\mu^{2}-1)\frac{\partial}{\partial\mu}{}_{s}Y_{\ell m}=\ell{}_{s}H_{(\ell+1)m}Y_{(\ell+1)m}-(\ell+1){}_{s}H_{\ell m}Y_{(\ell-1)m}+\frac{sm}{\ell(\ell+1)}{}_{s}Y_{\ell m}\nonumber \\
 &  & \mu{}_{s}Y_{\ell m}={}_{s}H_{(\ell+1)m}{}_{s}Y_{(\ell+1)m}{}_{s}+{}_{s}H_{\ell m}{}_{s}Y_{(\ell-1)m}-\frac{sm}{\ell(\ell+1)}{}_{s}Y_{\ell m}\,
\end{eqnarray}
where
\begin{equation}
{}_{s}H_{\ell m}\equiv \sqrt{\frac{(\ell^{2}-m^{2})(\ell^{2}-s^{2})}{\ell^{2}(4\ell^{2}-1)}}\,,
\end{equation}
with $s=\pm2$, and keeping just the leading terms in $\beta$, one finally finds
\begin{eqnarray}
    a_{\ell m}^{E\,\prime} & = & \sum_{\ell^{\prime}}\Big\{\delta_{\ell^{\prime}\ell} -\beta\Big[{}_{+2}H_{(\ell+1)m}(\ell-d+2)\delta_{\ell(\ell^{\prime}-1)}-{}_{+2}H_{\ell m}(\ell+d-1)\delta_{\ell(\ell^{\prime}+1)}\Big]\Big\}\, a_{\ell^{\prime}m}^{E}\nonumber \\
    &  & -i\,\beta\sum_{\ell^{\prime}}\frac{(2d-2)m}{\ell(\ell+1)} \delta_{\ell^{\prime}\ell}\, a_{\ell^{\prime}m}^{B}+\mathcal{O}(\beta^{2})\nonumber \\
    & \equiv & a_{\ell m}^{E}+\left(c_{\ell m}^{-E}a_{\ell-1m}^{E}+c_{\ell m}^{+E}a_{\ell+1m}^{E}\right)
\end{eqnarray}
and
\begin{eqnarray}
    a_{\ell m}^{B\,\prime} & = & \sum_{\ell^{\prime}}\Big\{\delta_{\ell^{\prime}\ell} -\beta\Big[{}_{+2}H_{(\ell+1)m}(\ell-d+2)\delta_{\ell(\ell^{\prime}-1)}-{}_{+2}H_{\ell m}(\ell+d-1)\delta_{\ell(\ell^{\prime}+1)}\Big]\Big\}\, a_{\ell^{\prime}m}^{B}\nonumber \\
    &  & +i\,\beta\sum_{\ell^{\prime}}\frac{(2d-2)m}{\ell(\ell+1)} \delta_{\ell^{\prime}\ell}a_{\ell^{\prime}m}^{E}+\mathcal{O}(\beta^{2})\nonumber \\
    & \equiv & a_{\ell m}^{B}+\left(c_{\ell m}^{-B}a_{\ell-1m}^{B}+c_{\ell m}^{+B}a_{\ell+1m}^{B}\right)\,,\label{aintEB}
\end{eqnarray}
where one sees that for the cross $E$, $B$ terms the inclusion of the Doppler effect ($d=1$) exactly cancels that of aberration, resulting in no correlation between both polarization modes. This should be contrasted with the results in~\cite{challinor02}, that found a nonzero correlation between $E$ and $B$ modes already at ${\cal O}(\beta)$. Here it exactly vanishes because we are considering quantities which scale under a Lorentz boost according to Eq.~(\ref{boost}) instead of according to~\eqref{I}.

\section{Analysis in cartesian components}
\label{app:gendir}

If we take a generic velocity, not necessarily oriented towards the $\hat{z}$ axis,
considering the $b_{lm}$ coefficients of eq.\eqref{almblm}, obtained at lowest order in $\beta$, and rotating them as in eq.\eqref{eq:alm-rotated} we get
\begin{equation}
b_{1s}\,=\, D^1_{s 0}(\phi, \theta, 0) \,b_{10}' \,=\, \sqrt{\frac{4\pi}{3}} Y^{*}_{1s}(\theta,\phi) \,b_{10}' \,=\, \frac{4\pi}{3}Y^{*}_{1s} (\hat{\boldsymbol \beta}) \beta \,\equiv\, \sqrt{\frac{4\pi}{3}} \beta_s\,,
\end{equation}
where $s=\{-1,0,+1\}$ and where in the last equality we defined the new variables $\{\beta_{-},\beta_{0},\beta_{+}\}$ for convenience. Note that from the definitions of $\beta_s$ one also has that the cartesian components of $\beta$ are just
\begin{equation} \label{eq:beta-x-y}
\begin{aligned}
    \beta_{x} \,\;&= -\sqrt{2} \, {\textrm{Re}}[\beta_+]  \,,\\
    \beta_{y} \,\;&=\, \;\; \sqrt{2} \,{\textrm{Im}}[\beta_+]  \,, \\
    \beta_{z} \,\;&=\, \;\;\beta_0 \,.
\end{aligned}
\end{equation}
Now, since
\begin{equation}
\label{eq:almL-gaunt-new}
    a_{\ell\, m}^{(L)}\simeq\sum_{\ell's}(-1)^{m} \,G_{\;\;\;\ell}^{-m}{\,}_{\;\;\;\ell'}^{m-s}{\,}_1^{s} \; \frac{\ell'(\ell'+1)-\ell(\ell+1)+2}{2}\;a_{\ell'\, m}^{(P)}\,b_{1s}\,,
\end{equation}
we can write
\begin{equation}
    a_{\ell\, m}^{(L)}\simeq \sum_{s=-1}^{1} \beta_s \left[ c_{\ell m}^{s-}a_{\ell-1\, m-s}^{(P)}+c_{\ell m}^{s+}a_{\ell+1\, m-s}^{(P)} \right],
    \label{basicROT}
\end{equation}
with the new coefficients given by
\begin{equation}
\begin{aligned}
    c_{\ell \,m}^{s\,+} & =  (-1)^{s} (\ell +2)\sqrt{\frac{(\ell +1+m-s)(\ell +1-m+s)\left(\ell -s m+s^2\right)}{\left(4(\ell +1)^2-1\right)\left(1+s^2\right)(\ell +s m)}}\,, \\
    c_{\ell \,m}^{s\,-} & =  - (\ell -1)\sqrt{\frac{(\ell +m-s)(\ell -m+s)\left(\ell +s m+1-s^2\right)}{\left(4\ell ^2-1\right)\left(1+s^2\right)(\ell -s m+1)}}\,,
    \label{newcoeff}
\end{aligned}
\end{equation}
Note that these coefficients exhibit the following interesting property, which can be easily verified by inspection:
\begin{align}
    c_{\ell \,m}^{s\,\pm}  = c_{\ell \,-m}^{-s\,\pm} \,.
\end{align}

In order to build the estimator let us define again (cf.~\eqref{eq:efdef-1})
\begin{align}
\label{eq:efdef}
    F_{\ell \, m}^s \equiv a_{\ell \, m}^*  a_{\ell+1 \, m+s}\,.
\end{align}
which satisfies $F_{\ell m}^{\pm *}=-F_{\ell -m}^{\mp}$.

Note that we have rotated eq.\eqref{basic} with the coefficients \eqref{cABERR} getting eq.\eqref{basicROT} with coefficients \eqref{newcoeff} for the aberration effect, finding that $s$ can be only $0,+1,-1$. Including also the Doppler effect only amounts to rotate with the coefficients \eqref{cTOTAL}, which only differ by a $d-$dependent prefactor, so that we get a similar result with the new coefficients:
\begin{equation}
\label{clmsT}
\begin{aligned}
    c_{\ell \,m}^{s\,+} & =  (-1)^{s} (\ell +2-d)\sqrt{\frac{(\ell +1+m-s)(\ell +1-m+s)\left(\ell -s m+s^2\right)}{\left(4(\ell +1)^2-1\right)\left(1+s^2\right)(\ell +s m)}}\,, \\
    c_{\ell \,m}^{s\,-} & =  - (\ell -1+d)\sqrt{\frac{(\ell +m-s)(\ell -m+s)\left(\ell +s m+1-s^2\right)}{\left(4\ell ^2-1\right)\left(1+s^2\right)(\ell -s m+1)}}\,,
\end{aligned}
\end{equation}

Now by applying~\eqref{basicROT} to~\eqref{eq:efdef}, we find that
\begin{align}
    \left<F_{\ell \, m}^s \right>  \,=\, \beta_s \left[c_{\ell+1 \, m+s}^{s\,-}\, C_\ell + (-1)^s  c_{\ell \, m}^{-s\,+} \,C_{\ell+1} \right]\,,
\end{align}
in which we used that $(\beta_s)^* = (-1)^s \beta_{-s}$ and the usual relation $\left<a_{\ell m}^* a_{\ell' m'} \right> = C_\ell \delta_{\ell \ell'}\delta_{m m'}$. This is a powerful result since each $\left<F_{\ell \, m}^s \right>$  depends only on a single $\beta_s$. This feature naturally implies that the Fisher Matrix is a diagonal matrix. For the 4-point correlation functions, assuming Gaussianity, we find that
\begin{align}
    \left<F_{\ell \, m}^s  \big(F_{\ell' \, m'}^{s'}\big)^* \right> - \Big<F_{\ell \, m}^s\Big>  \Big<\big(F_{\ell' \, m'}^{s'}\big)^* \Big> \,=\, C_\ell \,C_{\ell+1} \,\delta_{\ell \ell'} \delta_{m m'} \delta_{s s'}\,.
\end{align}

Since it is convenient to work with real quantities, we define the real and imaginary parts of $F_{\ell \, m}^s$ by:
\begin{equation}
\begin{aligned}
    f_{\ell \, m}^s &\,\equiv\, \frac{1}{2} \left[ F_{\ell \, m}^s  + \left(F_{\ell \, m}^s\right)^* \right]\,, \\
    g_{\ell \, m}^s &\,\equiv\, \frac{1}{2i} \left[ F_{\ell \, m}^s  - \left(F_{\ell \, m}^s\right)^* \right]\,.
\end{aligned}
\end{equation}
These quantities in turn satisfy the following:
\begin{equation}
\begin{aligned}
    \left<f_{\ell \, m}^s \right>  &\,=\, {\textrm{Re}}\left[\beta_s\right] \left[c_{\ell+1 \, m+s}^{s\,-}\, C_\ell + (-1)^s  c_{\ell \, m}^{-s\,+} \,C_{\ell+1} \right] \,, \\
    \left<g_{\ell \, m}^s \right>  &\,=\, {\textrm{Im}}\left[\beta_s\right] \left[c_{\ell+1 \, m+s}^{s\,-}\, C_\ell + (-1)^s  c_{\ell \, m}^{-s\,+} \,C_{\ell+1} \right] \,,
\end{aligned}
\end{equation}
and
\begin{equation}
\begin{aligned}
    \left<f_{\ell \, m}^s \, f_{\ell' \, m'}^{s'}\right> - \Big<f_{\ell \, m}^s\Big> \Big<f_{\ell' \, m'}^{s'}\Big> \,&=\, \frac{1}{2} C_\ell \,C_{\ell+1} \,\delta_{\ell \ell'} \left( \delta_{s s'}\delta_{m m'}+\delta_{s -s'}\delta_{m -m'} \right)\,, \\
    \left<g_{\ell \, m}^s \, g_{\ell' \, m'}^{s'}\right> - \Big<g_{\ell \, m}^s\Big> \Big<g_{\ell' \, m'}^{s'}\Big> \,&=\, \frac{1}{2} C_\ell \,C_{\ell+1} \,\delta_{\ell \ell'} \left( \delta_{s s'}\delta_{m m'}-\delta_{s -s'}\delta_{m -m'} \right)\,, \\
    \left<f_{\ell \, m}^s \, g_{\ell' \, m'}^{s'}\right> - \Big<f_{\ell \, m}^s\Big> \Big<g_{\ell' \, m'}^{s'}\Big> \,&=\, 0\,.
\end{aligned}
\end{equation}

A crucial property regarding these quantities is
\begin{equation}\label{eq:flms-glms-symmetries}
    \,f_{\ell \, m}^s = (-1)^s f_{\ell \, -m}^{-s}\,\qquad \mbox{and} \qquad \,g_{\ell \, m}^s = (-1)^{s+1} g_{\ell \, -m}^{-s}\,.
\end{equation}
This implies that all $f_{\ell \, m}^s$ and $g_{\ell \, m}^s$ for negative $m$ values can be written in terms of their counterparts with positive $m$'s, so that we must only consider, say, $m \ge 0$ when summing independent correlations. Nevertheless, even restraining to $m \ge 0$ one double-counts two quantities, to wit $f_{\ell \, 0}^{\pm 1}$ and $g_{\ell \, 0}^{\pm 1}$. Moreover, since $g_{\ell \, 0}^0$ is always zero, for a given $\ell$ one can have a total of $\,6 \ell + 3\,$ non-null independent quantities. Instead of restricting ourselves to $m \ge 0$ (and have to worry about $f_{\ell \, 0}^{\pm 1}$ and $g_{\ell \, 0}^{\pm 1}$) we shall count these independent quantities in the following way: $s \in \{0,+1\}$ and either $m \in \{-\ell,\ell\}$ (for $s=1$) or $m \in \{0,\ell\}$ (for $s=0$). With this convention the above relations simplifies to (using~\eqref{eq:beta-x-y})
\begin{equation}\label{eq:f-g-h-lms}
\begin{aligned}
    \left<f_{\ell \, m}^s \right>  &\,=\, \left[-\frac{\beta_x}{\sqrt{2}} \delta_{s1} + \beta_z \delta_{s0}  \right] h^s_{\ell m} \,, \\
    \left<g_{\ell \, m}^s \right>  &\,=\,\frac{\beta_y}{\sqrt{2}}  \delta_{s1} \, h^s_{\ell m} \,, \\
    h^s_{\ell m} &\,\equiv\, c_{\ell+1 \, m+s}^{s\,-}\, C_\ell + (-1)^s  c_{\ell \, m}^{-s\,+} \,C_{\ell+1}\,,
\end{aligned}
\end{equation}
and
\begin{equation}
\begin{aligned}
    \left<f_{\ell \, m}^s \, f_{\ell' \, m'}^{s'}\right> - \Big<f_{\ell \, m}^s\Big> \Big<f_{\ell' \, m'}^{s'}\Big> \,&=\, \frac{1}{2} C_\ell \,C_{\ell+1} \,\delta_{\ell \ell'} \delta_{s s'} \left( \delta_{m m'}+\delta_{s 0}\delta_{m0} \right)\,, \\
    \left<g_{\ell \, m}^s \, g_{\ell' \, m'}^{s'}\right> - \Big<g_{\ell \, m}^s\Big> \Big<g_{\ell' \, m'}^{s'}\Big> \,&=\, \frac{1}{2} C_\ell \,C_{\ell+1} \,\delta_{\ell \ell'} \delta_{s s'} \left(\delta_{m m'}-\delta_{s 0}\delta_{m0} \right)\,.
    \label{varianze}
\end{aligned}
\end{equation}

A similar estimator can be constructed in order to study the polarization. Following the same notation used in previous sections, we define
\begin{align}
    f_{\ell \, m}^{s X Y} \,\equiv\, \frac{1}{2} \left[ F_{\ell \, m}^{s X Y}  + \left(F_{\ell \, m}^{s X Y}\right)^* \right]\,, \\
    g_{\ell \, m}^{s X Y} \,\equiv\, \frac{1}{2i} \left[ F_{\ell \, m}^{s X Y}  - \left(F_{\ell \, m}^{s X Y}\right)^* \right]\,.
\end{align}
Applying then a Wigner rotation to Eq.~(\ref{eq:genform}), we find for such an estimator
\begin{equation}\label{eq:f-g-h-lms-pol}
\begin{aligned}
    \left<f_{\ell \, m}^{s X Y} \right>  &\,=\, \left[-\frac{\beta_x}{\sqrt{2}} \delta_{s1} + \beta_z \delta_{s0}  \right] h^{s X Y}_{\ell m} \,, \\
    \left<g_{\ell \, m}^{s X Y} \right>  &\,=\,\frac{\beta_y}{\sqrt{2}}  \delta_{s1} \, h^{s X Y}_{\ell m} \,, \\
    h^{s X Y}_{\ell m} &\,\equiv\, c_{\ell+1 \, m+s}^{s\,-\,Y}\, C_\ell^{X Y} + (-1)^s  c_{\ell \, m}^{-s\,+\,X} \,C^{X Y}_{\ell+1}\,,
\end{aligned}
\end{equation}
and
\begin{equation}
\begin{aligned}
    \left<f_{\ell \, m}^{s X Y} \, f_{\ell' \, m'}^{s' H K}\right> - \Big<f_{\ell \, m}^{s X Y}\Big> \Big<f_{\ell' \, m'}^{s' H K}\Big> \,&=\, \frac{1}{2} C_\ell^{X H} \,C_{\ell+1}^{Y K} \,\delta_{\ell \ell'} \delta_{s s'} \left( \delta_{m m'}+\delta_{s 0}\delta_{m0} \right)\,, \\
    \left<g_{\ell \, m}^{s X Y} \, g_{\ell' \, m'}^{s' H K}\right> - \Big<g_{\ell \, m}^{s X Y}\Big> \Big<g_{\ell' \, m'}^{s' H K}\Big> \,&=\, \frac{1}{2} C_\ell^{X H} \,C_{\ell+1}^{Y K} \,\delta_{\ell \ell'} \delta_{s s'} \left(\delta_{m m'}-\delta_{s 0}\delta_{m0} \right)\,,
    \label{varianze-pol}
\end{aligned}
\end{equation}
where the coefficients $c_{\ell \, m}^{-s\,\pm\,E,B}$ are given by
\begin{equation}
\begin{aligned}
    c_{\ell \,m}^{s\,\pm\,E,B} & = \sqrt{\frac{\ell^2-4}{\ell^2}} c_{\ell \,m}^{s\,\pm,T}
\end{aligned}
\end{equation}
and $c_{\ell \,m}^{s\,\pm\,T} = c_{\ell \,m}^{s\,\pm}$ (see Eq.~(\ref{clmsT})). Let us also note that for large $\ell$, $c_{\ell \, m}^{-s\,\pm\,E,B} \rightarrow c_{\ell \,m}^{s\,\pm}$. In particular, for $\ell = 15$ the difference is smaller than $1\%$.

%%%%%%%%%%%%%%%%%%%%%%%%%%%%%%%%%%%%%%%%%%%%%%%%%%%%%%%%%%%%%%%%%%%%%%%%%%%

\bibliography{arxiv-v3}

\end{document}